\documentclass[%
 aip,
 amsmath,amssymb,
 reprint,%
]{revtex4-1}

\usepackage{graphicx}
\usepackage{dcolumn}
\usepackage{bm}

\usepackage[utf8]{inputenc}
\usepackage[T1]{fontenc}
\usepackage{mathptmx}
\usepackage{stfloats}

\begin{document}

\preprint{APM19-AR-01221}

\title[Structural and electronic properties of the pure and stable elemental 3D TDS $\alpha$-Sn]{Structural and electronic properties of the pure and stable\\elemental 3D topological Dirac semimetal $\alpha$-Sn}

\author{Ivan Madarevic}
\author{Umamahesh Thupakula}%
\affiliation{ 
Quantum Solid State Physics, KU Leuven, Celestijnenlaan 200D, 3001 Leuven, Belgium
}
 \author{Gertjan Lippertz}
 \affiliation{ 
Quantum Solid State Physics, KU Leuven, Celestijnenlaan 200D, 3001 Leuven, Belgium
}
\affiliation{ 
Physics Institute II, University of Cologne, Zülpicher Str. 77, 50937 Köln, Germany
}
 \author{Niels Claessens}
 \affiliation{ 
Quantum Solid State Physics, KU Leuven, Celestijnenlaan 200D, 3001 Leuven, Belgium
}
\affiliation{ 
IMEC, Kapeldreef 75, 3001 Leuven, Belgium
}
 \author{Pin-Cheng Lin}
  \author{Harsh Bana}
  \affiliation{ 
Quantum Solid State Physics, KU Leuven, Celestijnenlaan 200D, 3001 Leuven, Belgium
}%

\author{Sara Gonzalez}
\author{Giovanni Di Santo}
\author{Luca Petaccia}
\affiliation{Elettra Sincrotrone Trieste, Strada Statale 14 km 163.5, 34149 Trieste, Italy
}%
\author{Maya Narayanan Nair}
\affiliation{ 
Quantum Solid State Physics, KU Leuven, Celestijnenlaan 200D, 3001 Leuven, Belgium
}%
\affiliation{ 
CUNY Advanced Science Research Center, 85 St. Nicholas Terrace, New York, NY 10031, USA
}
\author{Lino M.C. Pereira}
\author{Chris Van Haesendonck}
\author{Margriet J. Van Bael}
\affiliation{ 
Quantum Solid State Physics, KU Leuven, Celestijnenlaan 200D, 3001 Leuven, Belgium
}%

\date{\today}

\begin{abstract}
In-plane compressively strained $\alpha$-Sn films have been theoretically predicted and experimentally proven to possess non-trivial electronic states of a 3D topological Dirac semimetal. The robustness of these states typically strongly depends on purity, homogeneity and stability of the grown material itself. By developing a reliable fabrication process, we were able to grow pure strained \mbox{$\alpha$-Sn} films on InSb(100), without heating of the substrate during growth, nor using any dopants. The \mbox{$\alpha$-Sn} films were grown by molecular beam epitaxy, followed by experimental verification of the achieved chemical purity and structural properties of the film's surface. Local insight into the surface morphology was provided by scanning tunneling microscopy. We detected the existence of compressive strain using Mössbauer spectroscopy and we observed a remarkable robustness of the grown samples against ambient conditions. The topological character of the samples was confirmed by angle-resolved photoemission spectroscopy, revealing the Dirac cone of the topological surface state. Scanning tunneling spectroscopy, moreover, allowed obtaining an improved insight into the electronic structure of the 3D topological Dirac semimetal \mbox{$\alpha$-Sn} above the Fermi level.
\end{abstract}

\maketitle

%

\section{Introduction}

$\alpha$-Sn is a low-temperature phase of tin which in the bulk state becomes stable below 13.2 $^{\circ}$C. In contrast to its much more commonly known metallic allotrope $\beta$-Sn, bulk \mbox{$\alpha$-Sn} is a semi-metal (zero band gap) with a pronounced covalent character. This material has attracted the researchers’ attention since it has been theoretically predicted that \mbox{$\alpha$-Sn} is the only elemental strong 3D topological insulator (TI) if subjected to uniaxial strain, which would open a bulk band-gap \cite{Fu}. Strain in \mbox{$\alpha$-Sn} has been, to a certain extent, achieved in \mbox{$\alpha$-Sn} films, epitaxially grown on a substrate with appropriate lattice mismatch (InSb, CdTe) \cite{FARROW1981507,BETTI2002335,Yuen_1990}. Although the 2D electron gas behavior of the surface states of \mbox{$\alpha$-Sn} films was detected more than 30 years ago \cite{yuen1989first}, the topological character was confirmed only recently in Te and/or Bi doped thin films by angle-resolved photoemission spectroscopy (ARPES) measurements, depicting the surface electronic properties below the Fermi level \cite{Barfuss2013,Ohtsubo2013-alphaSnARPES,Xu2017-alphaSn111-semi,RogalevPhysRevB.95.161117, Scholz-PhysRevB.97.075101}. ARPES revealed a topological Dirac cone in the electronic band structure of doped \mbox{$\alpha$-Sn}, locating the Dirac point near the Fermi level. Interestingly, the most recent studies emphasized that the type of strain in the case of \mbox{$\alpha$-Sn} is important: in-plane compressive strain turns it into a TDS, while in-plane tensile strain transforms \mbox{$\alpha$-Sn} it into a TI \cite{Xu2017-alphaSn111-semi,RogalevPhysRevB.95.161117,Huang2017_tensileZZ,Scholz-PhysRevB.97.075101,Zhang_Luttinger_Sn2018, Rogalev2019}. Very recently, the signature of the topological charge carriers has also been demonstrated in electronic magneto-transport experiments through the analysis of Shubnikov-de Haas oscillations \cite{Barbedienne-PhysRevB.98.195445}. These interesting findings make \mbox{$\alpha$-Sn} films useful for spintronics and put this TDS in the group of materials which could become a building unit (qubit) of future quantum computers.

The growth of \mbox{$\alpha$-Sn} thin films is associated with difficulties, since this allotrope of Sn is not stable in bulk at room temperature. To stabilize it by inducing the epitaxial growth of \mbox{$\alpha$-Sn}, a suitable substrate with matching lattice constant is needed (acting as a crystallization seed). InSb(100) provides that matching and in addition introduces strain in the grown \mbox{$\alpha$-Sn} film. However, the use of InSb substrates brings along challenges such as surface degradation (InSb is very volatile to surface oxidation) and thermal instability (InSb decomposes far below its melting point), complicating the preparation of the substrate surface by annealing. These challenges can be, to a certain extent, overcome by employing a procedure which includes combined ion-beam sputtering and thermal annealing in ultra-high vacuum (UHV) \cite{FARROW1981507,Frost_1998-InSb_sput}. This procedure can lead to the formation of In islands on the substrate surface, enhancing the already known issue of In interdiffusion during deposition of Sn on InSb(100) substrates that are only slightly heated \cite{MAGNANO200229}. As a consequence, the surface analysis of \mbox{$\alpha$-Sn} films shows a substantial amount of In \cite{BETTI2002335,Ohtsubo2013-alphaSnARPES,Scholz-PhysRevB.97.075101}. At this point there is a gap in knowledge about the scale on which this affects the phase purity and consequentially, indirectly, the topological electronic transport properties of the \mbox{$\alpha$-Sn} films. Having this in mind, the most favorable scenario is to avoid In and Sb interdiffusion, while still achieving a good surface quality and with that maintaining the topological properties of \mbox{$\alpha$-Sn}.

As \mbox{$\alpha$-Sn} crystallizes in the diamond structure, the dangling $sp^{3}$-bonds of the top most surface atomic layer of the (100) planes are unsaturated, and they are pointing diagonally (lying in the (110) plane). Such bond orientation leads to the formation of asymmetric dimers on the film surface, similar to the cases of Si(100) and Ge(100) \cite{SemicondSurfaces-KAHN1983193,SiDimer1985PhysRevLett.55.1303,SiDimerTheoPhysRevB.51.14504,STMsemi-KUBBY199661}. In previous works, in order to avoid Sn dimer formation and thus enhance surface quality of the grown films, elevated substrate temperature and/or doping with substantial amounts of Bi \cite{Ohtsubo2013-alphaSnARPES,Barbedienne-PhysRevB.98.195445} and Te \cite{Barfuss2013,Scholz-PhysRevB.97.075101} were employed. The use of Bi and Te results in a disturbed chemical purity of the Sn film surface and a modified surface reconstruction. The influence that these impurities, combined with In interdiffusion, have on the stability and the topological electronic transport properties of \mbox{$\alpha$-Sn} is for now difficult to predict. Particularly intriguing can be the use of Te as a surfactant for the \mbox{$\alpha$-Sn} film growth, since the possible forming of Sn$_{1-x}$Te$_{x}$ would make the surface very susceptible to degradation \cite{SnTeOXY_BERCHENKO2018134,SnTe_degrad_LIU2019351} and would result in a slight distortion of the cubic crystal structure \cite{SnTeMOSS_FANO1977467,SnTe_Landolt1998}. If 3D TDSs are to be used as a robust unit for quantum computing related applications ("chiral qubits") \cite{li2020chiral, philip2020chiral}, growing a pure, continuous and homogeneous \mbox{$\alpha$-Sn} film surface is a priority.

Here we present the possibility of growing high-quality films of the 3D TDS \mbox{$\alpha$-Sn} in its full elemental form, showing a great stability. A combined study of growth, structural and electronic properties of \mbox{$\alpha$-Sn} is presented. We demonstrate the existence of strain with conversion electron Mössbauer spectroscopy (CEMS) and take a special closer look at the local surface morphology of the \mbox{$\alpha$-Sn} films with scanning tunneling microscopy (STM). We report topological surface electronic states in this type of samples using ARPES, and also provide a unique insight into the surface electronic properties above the Fermi level using scanning tunneling spectroscopy (STS).

\section{Sample fabrication}

Before the film growth, the high-quality single-crystal undoped (n-type) InSb(100) substrates (Wafer Technology Ltd.) were cleaned and polished in UHV environment using a multi-step process (see Supplementary Material). This included several cycles of Ar$^+$ ion-beam sputtering, using two energy modes in order to clean and achieve a gentle polishing of the InSb(100) substrate surface. Every cycle of the ion-beam sputtering is followed by a short multi-step thermal annealing cycle. The preparation process was monitored by Auger electron spectroscopy (AES), low-energy electron diffraction (LEED) and reflection high-energy electron diffraction (RHEED). This multi-step preparation procedure resulted in atomically flat and contamination free InSb(100) surfaces, with maintained stoichiometry and the characteristic c($8\times2$) surface reconstruction (see Supplementary Material). This achievement was crucial for the subsequent successful \mbox{$\alpha$-Sn} films growth.

The \mbox{$\alpha$-Sn} films were grown using molecular beam epitaxy (MBE), by evaporating Sn from a Knudsen cell onto a InSb(100) substrate held at 5$^{\circ}$C. Some of the samples were also grown on slightly heated substrates (30, 50 and 80$^{\circ}$C) for the sake of comparison (see Supplementary Material). Rutherford backscattering spectrometry (RBS) was used on calibration samples in order to precisely determine the Sn deposition rate. In this way, a film thickness accuracy in the range of \mbox{$\pm1$ nm} was achieved. Sn films of 10, 20 and \mbox{30 nm} thickness were successfully grown while being monitored using RHEED. During the Sn deposition, RHEED oscillations were recorded (Fig.~\ref{fig1}(a)) and it was confirmed that the growth rate ($10$ \AA/min) is matching the targeted deposition rate.

\section{Results and discussion}

\subsection{Chemical composition and surface structure}

\begin{figure*}
\includegraphics[width=\linewidth,height=\textheight,keepaspectratio]{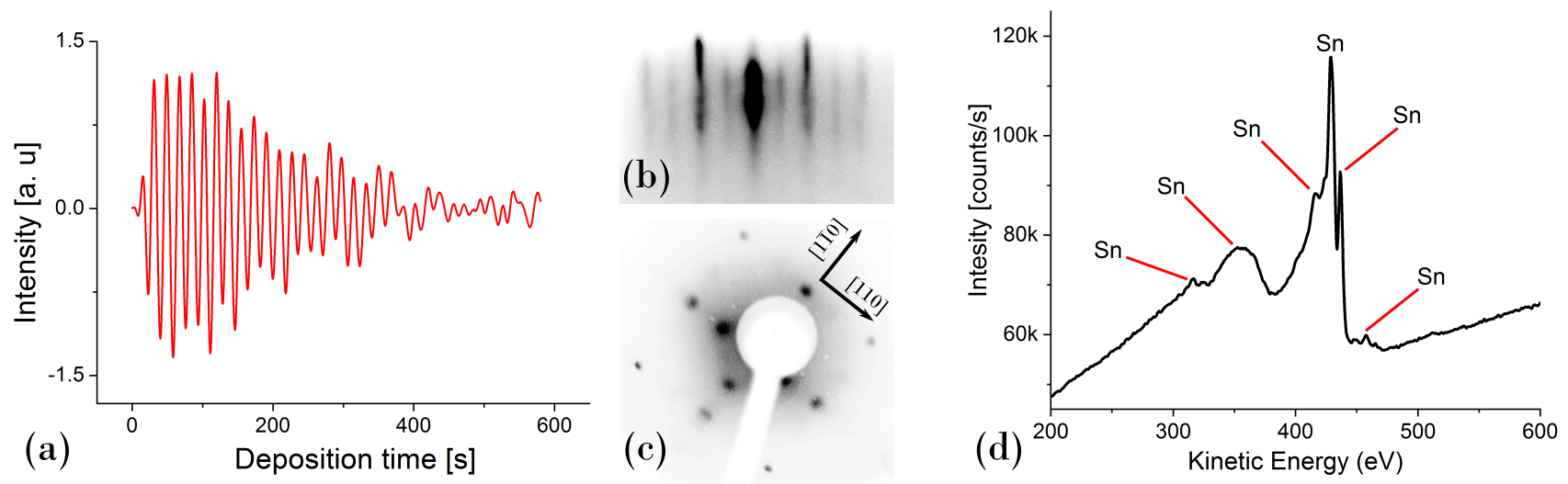}
\caption{\label{fig1} Growth data for a \mbox{20 nm} thick \mbox{$\alpha$-Sn} film grown at a substrate temperature of \mbox{5$^{\circ}$C}. (a) RHEED oscillations (with high frequencies filtered out); 
 (b) RHEED (\mbox{10 kV}) image taken along the [110] direction, showing two-domain (\mbox{2$\times$1}) pattern;
 (c) LEED (\mbox{48 eV}) image of two-domain (\mbox{2$\times$1}) pattern;
 (d) AES spectrum taken in situ after deposition, showing only the MNN spectral lines of Sn.}
 \end{figure*}

The chemical purity of the surface in the case of 20 and \mbox{30 nm} thick Sn films grown at a substrate temperature of 5$^{\circ}$C is confirmed by AES (Fig.~\ref{fig1}(d)). The films are uniformly covered with Sn with no traces of other chemical elements, indicating an excellent elemental purity of the Sn film surface. The AES spectra of the thinner films (\mbox{10 nm}) show minor traces of In, while in the cases where the InSb(100) substrate was heated In and Sb were also detected by AES, regardless of the thickness (see Supplementary Material). This implies that a slight increase in the substrate temperature (around room temperature) deteriorates the chemical homogeneity of the surface. The X-ray photoelectron spectroscopy (XPS) spectra of the grown films do show (see Supplementary Material) a very small presence of In (\mbox{< 2 \%} and \mbox{< 1 \%} for 20 and \mbox{30 nm} thick Sn films, respectively), originating from the deeper layers of the sample, since the AES spectra show no traces of In in these cases. This is only partially in agreement with previous reports about \mbox{$\alpha$-Sn/InSb} pseudomorphic growth \cite{BETTI2002335,MAGNANO200229}. Indium interdiffusion is indeed enhanced if the temperature of the substrate is elevated. However, contrary to these reports we were able to observe "layer by layer" growth (Fig.~\ref{fig1}(a)) without raising the temperature of the InSb(100) substrate, and maintain the full elemental purity of the Sn film surface. RHEED and LEED images (Fig.~\ref{fig1}(b) and (c)) exhibit sharp features of the two-domain ($2\times1$) surface reconstruction of \mbox{$\alpha$-Sn}.

\begin{figure*}
\includegraphics[width=\textwidth,height=\textheight,keepaspectratio]{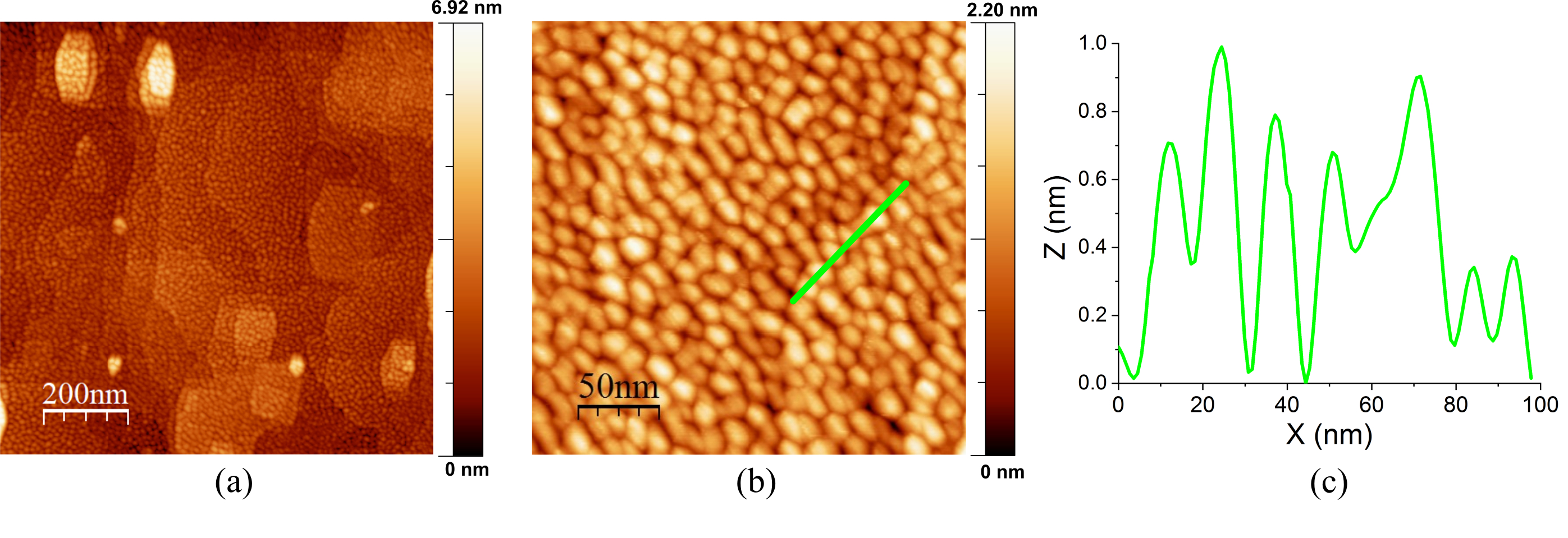}
 \caption{\label{fig2} STM (room temperature) topographic images of a \mbox{20 nm} thick \mbox{$\alpha$-Sn} film grown at a substrate temperature of \mbox{5$^{\circ}$C}. (a) \mbox{500 nm} $\times$ \mbox{500 nm} (\mbox{$U = 1.5$ V}, \mbox{$I = 100$ pA}); (b) \mbox{260 nm} $\times$ \mbox{260 nm} (\mbox{$U = 1.5$ V}, \mbox{$I = 140$ pA}); (c) Line profile along the green line in the panel (b).}
 \end{figure*}

The STM topography images (Fig.~\ref{fig2}) acquired in situ at room temperature show the same features for all \mbox{$\alpha$-Sn} samples grown at a substrate temperature of 5$^{\circ}$C. Uniform surfaces with sub-nanometer roughness were observed (\mbox{$RMS_{500}\sim0.4\; \mathrm{nm}$} -- \textit{root mean square roughness} for a $500 \times 500\; \mathrm{nm}^2$ surface). The terraces of the substrate are covered with evenly spread grains, still leaving the substrate steps visible in the overlay (Fig.~\ref{fig2}(a)). The average lateral size of the grains is $10 - 20\; \mathrm{nm}$. Linking this finding to the electron diffraction data, we confirm a high degree of crystallinity of these grains which, combined with the low film roughness, accounts for the acquired sharp RHEED and LEED patterns.

In contrast to the shown LEED, RHEED and STM data (for non-heated substrates), the films grown on slightly heated InSb(100) exhibited disturbed epitaxial growth (see Supplementary Material).

Figure~\ref{fig3} presents the analysis of the CEMS data of a \mbox{20 nm} thick \mbox{$\alpha$-Sn} film. The acquired spectra are almost completely dominated by the \mbox{$\alpha$-Sn} peak at \mbox{2.13(1) mm/s}, considering that the CEMS sensitivity for $\mathrm{SnO}_2$ (small peak observed at \mbox{0 mm/s}) is approximately ten times larger compared to that for \mbox{$\alpha$-Sn} and \mbox{$\beta$-Sn} \cite{Greenwood1971}. Importantly, the peak from \mbox{$\beta$-Sn} is completely absent (appearing at \mbox{2.64(1) mm/s} for our Sn/SiO$_2$/Si reference samples (see Supplementary Material)). The Mössbauer isomer shift ($\delta$) is defined in the hyperfine interactions theory \cite{Greenwood1971} as a difference between the electron density values near the nucleus of the absorber (sample), $\sim|\Psi_A (0)| ^2$, and the source of the probing \mbox{$\gamma$-photon}, $\sim|\Psi_S (0)| ^2$, with
\begin{equation}
    \delta = constant\times \left \{ |\Psi_A (0)| ^2 - |\Psi_S (0)| ^2 \right \}.
\end{equation}
The significantly higher $\delta$ value for \mbox{$\alpha$-Sn} in our samples, compared to bulk samples \cite{Greenwood1971,Stevens1983,Svane1997,Sn-bulk_SKWAREK201610,Sn-bulk_ZACHARIASZ2018165} and films \cite{Gomez_CdTe_PhysRevB.67.115340,Kelly-PhysRevB.100.075408}, implies an elevated 5$s$ electron density value \cite{Greenwood1967symp,GREENWOOD1968339} as this shift cannot be an intrinsic effect of the Sn nucleus, but rather results from a difference in the electron density around it. Since we observed only one crystallographic site of Sn, without any quadruple splitting, and since the crystal structure of the grown \mbox{$\alpha$-Sn} films remains unchanged compared to bulk \mbox{$\alpha$-Sn}, the existence of the compressive strain is confirmed \cite{Unzueta2018}. This strain indirectly diminishes the screening effect of the $5p$ electrons towards the $5s$ electrons by enhancing the covalent nature of the bonding in strained \mbox{$\alpha$-Sn}.

Unfortunately, by using the CEMS technique alone, it is not possible to estimate the magnitude of strain in this particular case. The energy resolution of Mössbauer spectroscopy is, without doubt, astonishing (in the range of neV), but being a relative technique, to estimate the magnitude of strain a comparative study involving X-ray diffraction spectroscopy (XRD) and/or first-principles calculations is needed. Conclusion drawn from the CEMS data is in accordance with the experimentally determined room temperature bulk lattice parameter values for \mbox{$\alpha$-Sn} and InSb reported in the literature (\mbox{$a \geqslant 6.489$ \AA} for \mbox{$\alpha$-Sn} \cite{Alpha_latticeBROWNLEE1950, Alpha_lattice1954,wyckoff1963crystal}, and \mbox{$a \leqslant 6.479$ \AA} in the case of InSb \cite{Liu_InSbLAT1951,willardson1968InSbLAT,Adachi_InSbLAT1999,InSb_LAT_III-Vbook}). Moreover, authors who have grown thicker films were able to directly measure this difference in lattice parameters \cite{FARROW1981507,AsomFarrow1989} again showing that the lattice parameter of the room temperature \mbox{$\alpha$-Sn} is indeed slightly bigger than in the case of InSb. In order to estimate the magnitude of strain in our samples, we have also performed XRD \mbox{($\theta$-$2\theta$)} measurements. We have detected no lattice relaxation in the grown \mbox{$\alpha$-Sn} films (20 and 30 nm) and estimated the lattice parameter of the prepared InSb substrate (after sputtering and annealing) to be \mbox{6.475 \AA}. Comparing this value with the value for \mbox{$\alpha$-Sn} \mbox{6.489 \AA}, we can conclude that the compressive strain of \mbox{$\sim 0.22$ \%} corresponds to the detected increase of the IS value of 0.13 mm/s measured by CEMS.

Excellent crystallinity and phase purity of the \mbox{$\alpha$-Sn} samples in this study are crucial for the success of the further ex-situ measurements (ARPES and STS). The robustness of the samples allowed using Ar$^+$ sputtering to clean the surface before these measurements. Even if the samples were kept under normal ambient conditions for several weeks, it was still  possible to remove contamination by low-energy Ar$^+$ sputtering, observing only Sn in the AES spectrum and fully recovering the two-domain ($2\times1$) reconstruction of \mbox{$\alpha$-Sn}, as reflected by LEED and RHEED (see Supplementary Material). This implies prominent stability and robustness of the sample surface, which seemingly becomes passivated in air, not leading to further degradation of the grown film. Combined with the previous reports on the enhanced temperature stability of strained \mbox{$\alpha$-Sn} films (up to 170$^{\circ}$C) \cite{FARROW1981507,vsTEMP_Reno1989,vsTEMP_Song2019}, our findings give this material an additional application value. Having the reports on the fragile air-stability of the more widely investigated 3D topological materials Bi$_2$Te$_3$ \cite{BiTe1_Bando_2000, BiTe2_Netsou_2017,BiTe3_Music_2017} and Bi$_2$Se$_3$ \cite{BiSe1_Kong, BiSe2_Edmonds} (caused by adsorption of molecular species from ambient), this finding could make \mbox{$\alpha$-Sn} more favorable for applications.

\subsection{Surface electronic structure}

After growth, \mbox{$\alpha$-Sn} film samples were exposed to ambient conditions and transferred to the ARPES setup of the BaDElPh beamline \cite{PETACCIA_Elettra2009780} at the Elettra synchrotron. Before the measurement, the surface of the films was cleaned by Ar$^+$ sputtering (500 eV). The conducted ARPES measurements on \mbox{20 nm} thick \mbox{$\alpha$-Sn}, grown at a substrate temperature of 5$^{\circ}$C, depict the topologically protected surface electronic states, forming a Dirac cone at the $\overline{\Gamma}$ point (Fig.~\ref{fig4}(a)) as a consequence of the lack of back-scattering of the topological electronic charge carriers. This proves that topological surface states are indeed an intrinsic property of pure elemental \mbox{$\alpha$-Sn} \cite{Fu}, and can also be detected in samples grown without the use of any Bi or Te dopants. The intensity of the background overlay around the zero energy value (Fig.~\ref{fig4}(a)) shows, as previously reported \cite{Barfuss2013,Scholz-PhysRevB.97.075101}, that the $\Gamma_8^+$ bulk energy band crosses the Fermi level ($E_F$). By fitting the line profiles of the acquired spectrum (Fig.~\ref{fig4}(b)) we were able to estimate the position of the Dirac point ($E_D$) at $\sim 60 \textnormal{ meV}$ above $E_F$ (slight p-type behavior). From the fit of the slopes of the Dirac cone branches we extracted Fermi velocity values ($v_F$) of $0.48(1)\times10^6 \textnormal{ m/s}$ (left) and $0.52(1)\times10^6 \textnormal{ m/s}$ (right). These $v_F$ values are close to the theoretically predicted value for thick films of elemental \mbox{$\alpha$-Sn} ($0.58\times10^6 \textnormal{ m/s}$) \cite{Kufner2014AlhpaSn-topVS}, but clearly reduced compared to the previously reported values for Bi/Te doped \mbox{$\alpha$-Sn} ($\sim0.7\times10^6 \textnormal{ m/s}$) \cite{Ohtsubo2013-alphaSnARPES,Scholz-PhysRevB.97.075101}. This suggests that the effect of shifting $E_F$ upwards in the case of Bi/Te doped \mbox{$\alpha$-Sn} is actually accompanied by a contraction of the $\Gamma_7^-$ energy band (which hosts the Dirac cone), spanning, in that case, a smaller k-space in the surface Brillouin zone. To precisely determine if this is a direct effect of doping or rather an indirect effect induced by the change of strain in the grown films, additional and more detailed studies are needed.

As in the case of the ARPES experiments, the same type of samples were, after growth, transferred in nitrogen atmosphere to the low-temperature STM/STS setup (\mbox{4.5 K}). The samples were carefully cleaned by Ar$^+$ sputtering and a stable tunneling current was achieved. The acquired STM topography images (Fig.~\ref{fig4}(d)) showed the same features as in the case of our in situ STM measurements performed at room temperature (Fig.~\ref{fig2}). The measured differential tunneling conductance $dI/dV$ (area averaged) on top of the \mbox{$\alpha$-Sn} grains, shows a \mbox{p-type} behavior ($E_F$ shifted towards the valence band) of the grown film (Fig.~\ref{fig4}(c)), while deviations from this trend are noticed along the edges.  

The STS technique, in contrast to ARPES, can also probe the electronic states above the Fermi level. Hence it was possible to observe the existence of the surface state at the positive bias voltages in the acquired STS characteristics of our \mbox{$\alpha$-Sn} samples (Fig.~\ref{fig4}(c)). Additionally, comparing the differential tunneling conductance ($dI/dV$) spectrum with that of the ARPES, allows us to extract the topological surface state Dirac point of the \mbox{$\alpha$-Sn} film. The $dI/dV$ spectrum, which has been averaged over several \mbox{$\alpha$-Sn} grains, reveals two minima positions at the positive voltages, located at $\sim 70 \textnormal{ meV}$ and $\sim 280 \textnormal{ meV}$. Moreover, the $dI/dV$ spectrum also reveals that the Fermi level lies within the valence band, indicating the p-type behavior of our \mbox{$\alpha$-Sn} film akin to the ARPES observations. Comparing this to our ARPES data and having the previous ARPES reports on \mbox{$\alpha$-Sn/InSb} in mind \cite{Barfuss2013,Ohtsubo2013-alphaSnARPES,Scholz-PhysRevB.97.075101,Barbedienne-PhysRevB.98.195445,Rojas_PhysRevLett.116.096602,RogalevPhysRevB.95.161117}, we anticipate the first $dI/dV$ minimum ($\sim 70 \textnormal{ meV}$) to be the Dirac point of the topological Dirac cone of \mbox{$\alpha$-Sn/InSb} that exists above $E_F$ in our films, and the second minimum ($\sim 280 \textnormal{ meV}$) to be the bulk band edge contribution. On the other hand, if we analyze the spectra as in the case of Bi$_2$Se$_3$ (a 3D TI with similar electronic properties around $E_F$) \cite{BiSe1_Eelbo_2013STS,BiSe2_Fedotov_doi:10.1002/pssr.201800617}, we are able to estimate $E_D$ using the global minimum of the STS averaged spectrum (Fig.~\ref{fig4}(c)) around \mbox{70 meV} above $E_F$, where the bulk valence bands start to dominate. Such an electronic behavior is consistent with our ARPES measurements (the acquired $E_D$ values and the observations of the $\Gamma_8^+$ bulk energy band crossing $E_F$). It is also noteworthy that if single-point STS spectra are taken, the $E_D$ value fluctuates depending on the acquisition position (within the range \mbox{$50 - 100$ meV} above $E_F$). This is, in general, an expected nature of the STS spectra taken near the grain boundaries. Nevertheless, future grain size and grain shape dependent studies may clarify this electronic behavior.

Finally, we further comment on the observed p-type electrical behavior in our ARPES and STS data (Fig.~\ref{fig4}), which confirm the previous transport measurement reports \cite{FARROW1981507,Barbedienne-PhysRevB.98.195445}. Three possible causes for this electrical behavior of \mbox{$\alpha$-Sn} films have been suggested \cite{FARROW1981507}: thermal decomposition of the boron-nitride in the walls of the Sn-source Knudsen cell, autodoping through interdiffusion of the substrate elements, and a possible localized \mbox{$\beta$-Sn} nucleation. At this point we cannot exclude the boron impurities scenario (small amounts of boron are very difficult to detect by AES and XPS). It is also not possible to fully exclude the influence of In, whose signal was detected by XPS (absent in the AES spectrum). On the other hand, since no presence of the \mbox{$\beta$-Sn} nucleation was detected on the surface of our samples grown at a substrate temperature of 5$^{\circ}$C, the third scenario for explaining the origin of the p-type behavior is unlikely.

\section{Conclusion}

We successfully grew pure, strained 3D TDS \mbox{$\alpha$-Sn} films on InSb(100). To achieve this, two crucial steps were required: modifying the substrate preparation procedure and lowering the temperature of the substrate during the film growth. The quality of the samples was examined using multiple complementary techniques, including STM, providing new insights into the \mbox{$\alpha$-Sn} surface morphology. For the first time we report experimental evidence for the presence of compressive strain in \mbox{$\alpha$-Sn} films by Mössbauer spectroscopy. We observed an excellent robustness of the film surface against ambient conditions, which could be crucial for possible future applications. We verified the existence of the topological electronic states in \mbox{$\alpha$-Sn} using ARPES, and we characterized its electronic properties both below and above the Fermi level, using the STS technique.

\section{Experimental methods}

The MBE growth of the samples was performed by evaporating Sn (deposition/growth rate $\approx10$ \AA/min) from a Knudsen cell (1170$^{\circ}$C) in UHV ($<5 \times 10^{-10}\; \mathrm{mbar}$).

AES was conducted using an 8 kV electron beam while XPS was realized with the use of the \mbox{Mg K$\alpha$} X-rays ($E = 1.254\; \mathrm{keV}$).

RHEED and LEED images were recorded using an electron beam energy of 10 keV and 48 eV, respectively. Screen voltage used for LEED measurements was \mbox{1 keV}.

In situ room temperature STM was performed using an Omicron Nanotechnology LS-STM setup (base pressure of $10^{-10}\; \mathrm{mbar}$).

The CEMS experiments were carried out at room temperature in a low-pressure acetone gas atmosphere, using a parallel plate detector and a $^{119m} \mathrm{Sn/CaSnO}_{3}$ source. The spectra were analyzed relative to $\mathrm{SnO}_2$ using the "VindaD" package \cite{Gunnlaugsson2016}.

XRD measurements were performed using an X'Pert  'PANalytical' X-ray diffractometer using Cu $K_{\alpha 1}$ radiation

The ARPES experiments (21 eV, 77 K) were performed at BaDElPh beamline \cite{PETACCIA_Elettra2009780} at the Elettra synchrotron radiation facility in Trieste (Italy).

Ex situ STM and STS were performed using an Omicron Nanotechnology LT setup operated at 4.5 K (base pressure of $10^{-11}\; \mathrm{mbar}$). All of the STS $dI/dV$ spectra were acquired using a sample bias voltage \mbox{$U = 0.5$ V} and a tunneling current \mbox{$I = 300$ pA}. The spectrum was averaged over \mbox{5 nm} $\times$ \mbox{5 nm} areas on top of the \mbox{$\alpha$-Sn} grains and averaged over 6 grains.

\section{Supplementary material}

See supplementary material for the detailed information about the following: the InSb(100) substrate preparation process, XPS, LEED, RHEED and STM investigations of the \mbox{$\alpha$-Sn} films grown on slightly heated InSb(100) substrates, the CEMS data of \mbox{$\beta$-Sn} grown on \mbox{SiO$_2$/Si} substrate and the \mbox{Ar$^+$} cleaning procedure for the \mbox{$\alpha$-Sn} films grown at a substrate temperature of 5$^{\circ}$C.

\section{Acknowledgements}

The authors thank B. Opperdoes, L. Sancin and V. Joly for technical support. This work was supported by the Research Foundation -- Flanders (FWO) and by the KU Leuven C1 program grants No. C14/18/074 and  C12/18/006. G. L. acknowledges the support by the FWO, file No. 27531 and 52751. M. N. N. acknowledges funding from Horizon 2020-MSCA-IF-GA No. 796940. The authors wish to thank Elettra Sincrotrone Trieste for providing access to its synchrotron radiation facilities.

\section*{References}
\bibliography{aipsamp}

\begin{thebibliography}{57}%
\makeatletter
\providecommand \@ifxundefined [1]{%
 \@ifx{#1\undefined}
}%
\providecommand \@ifnum [1]{%
 \ifnum #1\expandafter \@firstoftwo
 \else \expandafter \@secondoftwo
 \fi
}%
\providecommand \@ifx [1]{%
 \ifx #1\expandafter \@firstoftwo
 \else \expandafter \@secondoftwo
 \fi
}%
\providecommand \natexlab [1]{#1}%
\providecommand \enquote  [1]{``#1''}%
\providecommand \bibnamefont  [1]{#1}%
\providecommand \bibfnamefont [1]{#1}%
\providecommand \citenamefont [1]{#1}%
\providecommand \href@noop [0]{\@secondoftwo}%
\providecommand \href [0]{\begingroup \@sanitize@url \@href}%
\providecommand \@href[1]{\@@startlink{#1}\@@href}%
\providecommand \@@href[1]{\endgroup#1\@@endlink}%
\providecommand \@sanitize@url [0]{\catcode `\\12\catcode `\$12\catcode
  `\&12\catcode `\#12\catcode `\^12\catcode `\_12\catcode `\%12\relax}%
\providecommand \@@startlink[1]{}%
\providecommand \@@endlink[0]{}%
\providecommand \url  [0]{\begingroup\@sanitize@url \@url }%
\providecommand \@url [1]{\endgroup\@href {#1}{\urlprefix }}%
\providecommand \urlprefix  [0]{URL }%
\providecommand \Eprint [0]{\href }%
\providecommand \doibase [0]{http://dx.doi.org/}%
\providecommand \selectlanguage [0]{\@gobble}%
\providecommand \bibinfo  [0]{\@secondoftwo}%
\providecommand \bibfield  [0]{\@secondoftwo}%
\providecommand \translation [1]{[#1]}%
\providecommand \BibitemOpen [0]{}%
\providecommand \bibitemStop [0]{}%
\providecommand \bibitemNoStop [0]{.\EOS\space}%
\providecommand \EOS [0]{\spacefactor3000\relax}%
\providecommand \BibitemShut  [1]{\csname bibitem#1\endcsname}%
\let\auto@bib@innerbib\@empty
\bibitem [{\citenamefont {Fu}\ and\ \citenamefont {Kane}(2007)}]{Fu}%
  \BibitemOpen
  \bibfield  {author} {\bibinfo {author} {\bibfnamefont {L.}~\bibnamefont
  {Fu}}\ and\ \bibinfo {author} {\bibfnamefont {C.~L.}\ \bibnamefont {Kane}},\
  }\bibfield  {title} {\enquote {\bibinfo {title} {Topological insulators with
  inversion symmetry},}\ }\href {\doibase 10.1103/PhysRevB.76.045302}
  {\bibfield  {journal} {\bibinfo  {journal} {Phys. Rev. B}\ }\textbf {\bibinfo
  {volume} {76}},\ \bibinfo {pages} {045302} (\bibinfo {year}
  {2007})}\BibitemShut {NoStop}%
\bibitem [{\citenamefont {Farrow}\ \emph {et~al.}(1981)\citenamefont {Farrow},
  \citenamefont {Robertson}, \citenamefont {Williams}, \citenamefont {Cullis},
  \citenamefont {Jones}, \citenamefont {Young},\ and\ \citenamefont
  {Dennis}}]{FARROW1981507}%
  \BibitemOpen
  \bibfield  {author} {\bibinfo {author} {\bibfnamefont {R.}~\bibnamefont
  {Farrow}}, \bibinfo {author} {\bibfnamefont {D.}~\bibnamefont {Robertson}},
  \bibinfo {author} {\bibfnamefont {G.}~\bibnamefont {Williams}}, \bibinfo
  {author} {\bibfnamefont {A.}~\bibnamefont {Cullis}}, \bibinfo {author}
  {\bibfnamefont {G.}~\bibnamefont {Jones}}, \bibinfo {author} {\bibfnamefont
  {I.}~\bibnamefont {Young}}, \ and\ \bibinfo {author} {\bibfnamefont
  {P.}~\bibnamefont {Dennis}},\ }\bibfield  {title} {\enquote {\bibinfo {title}
  {The growth of metastable, heteroepitaxial films of \mbox{$\alpha$-Sn} by
  metal beam epitaxy},}\ }\href {\doibase
  https://doi.org/10.1016/0022-0248(81)90506-6} {\bibfield  {journal} {\bibinfo
   {journal} {Journal of Crystal Growth}\ }\textbf {\bibinfo {volume} {54}},\
  \bibinfo {pages} {507--518} (\bibinfo {year} {1981})}\BibitemShut {NoStop}%
\bibitem [{\citenamefont {Betti}\ \emph {et~al.}(2002)\citenamefont {Betti},
  \citenamefont {Magnano}, \citenamefont {Sancrotti}, \citenamefont {Borgatti},
  \citenamefont {Felici}, \citenamefont {Mariani},\ and\ \citenamefont
  {Sauvage-Simkin}}]{BETTI2002335}%
  \BibitemOpen
  \bibfield  {author} {\bibinfo {author} {\bibfnamefont {M.~G.}\ \bibnamefont
  {Betti}}, \bibinfo {author} {\bibfnamefont {E.}~\bibnamefont {Magnano}},
  \bibinfo {author} {\bibfnamefont {M.}~\bibnamefont {Sancrotti}}, \bibinfo
  {author} {\bibfnamefont {F.}~\bibnamefont {Borgatti}}, \bibinfo {author}
  {\bibfnamefont {R.}~\bibnamefont {Felici}}, \bibinfo {author} {\bibfnamefont
  {C.}~\bibnamefont {Mariani}}, \ and\ \bibinfo {author} {\bibfnamefont
  {M.}~\bibnamefont {Sauvage-Simkin}},\ }\bibfield  {title} {\enquote {\bibinfo
  {title} {Growth morphology of ($1\times2$) \mbox{$\alpha$-Sn(100)}: a surface
  diffraction study},}\ }\href {\doibase
  https://doi.org/10.1016/S0039-6028(02)01267-0} {\bibfield  {journal}
  {\bibinfo  {journal} {Surface Science}\ }\textbf {\bibinfo {volume}
  {507-510}},\ \bibinfo {pages} {335--339} (\bibinfo {year}
  {2002})}\BibitemShut {NoStop}%
\bibitem [{\citenamefont {Yuen}\ \emph {et~al.}(1990)\citenamefont {Yuen},
  \citenamefont {Liu}, \citenamefont {Joyce},\ and\ \citenamefont
  {Stradling}}]{Yuen_1990}%
  \BibitemOpen
  \bibfield  {author} {\bibinfo {author} {\bibfnamefont {W.~T.}\ \bibnamefont
  {Yuen}}, \bibinfo {author} {\bibfnamefont {W.~K.}\ \bibnamefont {Liu}},
  \bibinfo {author} {\bibfnamefont {B.~A.}\ \bibnamefont {Joyce}}, \ and\
  \bibinfo {author} {\bibfnamefont {R.~A.}\ \bibnamefont {Stradling}},\
  }\bibfield  {title} {\enquote {\bibinfo {title} {{RHEED} studies of the
  surface morphology of \mbox{$\alpha$-Sn} pseudomorphically grown on
  {InSb}(100) by {MBE}-a new kind of non-polar/polar system},}\ }\href
  {\doibase 10.1088/0268-1242/5/5/001} {\bibfield  {journal} {\bibinfo
  {journal} {Semiconductor Science and Technology}\ }\textbf {\bibinfo {volume}
  {5}},\ \bibinfo {pages} {373} (\bibinfo {year} {1990})}\BibitemShut {NoStop}%
\bibitem [{\citenamefont {Yuen}\ \emph {et~al.}(1989)\citenamefont {Yuen},
  \citenamefont {Liu}, \citenamefont {Holmes},\ and\ \citenamefont
  {Stradling}}]{yuen1989first}%
  \BibitemOpen
  \bibfield  {author} {\bibinfo {author} {\bibfnamefont {W.}~\bibnamefont
  {Yuen}}, \bibinfo {author} {\bibfnamefont {W.}~\bibnamefont {Liu}}, \bibinfo
  {author} {\bibfnamefont {S.}~\bibnamefont {Holmes}}, \ and\ \bibinfo {author}
  {\bibfnamefont {R.}~\bibnamefont {Stradling}},\ }\bibfield  {title} {\enquote
  {\bibinfo {title} {First observation of a two-dimensional electron gas at the
  interface of \mbox{$\alpha$-Sn/{InSb} (100)} grown by molecular beam
  epitaxy},}\ }\href@noop {} {\bibfield  {journal} {\bibinfo  {journal}
  {Semiconductor Science and Technology}\ }\textbf {\bibinfo {volume} {4}},\
  \bibinfo {pages} {819} (\bibinfo {year} {1989})}\BibitemShut {NoStop}%
\bibitem [{\citenamefont {Barfuss}\ \emph {et~al.}(2013)\citenamefont
  {Barfuss}, \citenamefont {Dudy}, \citenamefont {Scholz}, \citenamefont
  {Roth}, \citenamefont {H\"opfner}, \citenamefont {Blumenstein}, \citenamefont
  {Landolt}, \citenamefont {Dil}, \citenamefont {Plumb}, \citenamefont
  {Radovic}, \citenamefont {Bostwick}, \citenamefont {Rotenberg}, \citenamefont
  {Fleszar}, \citenamefont {Bihlmayer}, \citenamefont {Wortmann}, \citenamefont
  {Li}, \citenamefont {Hanke}, \citenamefont {Claessen},\ and\ \citenamefont
  {Sch\"afer}}]{Barfuss2013}%
  \BibitemOpen
  \bibfield  {author} {\bibinfo {author} {\bibfnamefont {A.}~\bibnamefont
  {Barfuss}}, \bibinfo {author} {\bibfnamefont {L.}~\bibnamefont {Dudy}},
  \bibinfo {author} {\bibfnamefont {M.~R.}\ \bibnamefont {Scholz}}, \bibinfo
  {author} {\bibfnamefont {H.}~\bibnamefont {Roth}}, \bibinfo {author}
  {\bibfnamefont {P.}~\bibnamefont {H\"opfner}}, \bibinfo {author}
  {\bibfnamefont {C.}~\bibnamefont {Blumenstein}}, \bibinfo {author}
  {\bibfnamefont {G.}~\bibnamefont {Landolt}}, \bibinfo {author} {\bibfnamefont
  {J.~H.}\ \bibnamefont {Dil}}, \bibinfo {author} {\bibfnamefont {N.~C.}\
  \bibnamefont {Plumb}}, \bibinfo {author} {\bibfnamefont {M.}~\bibnamefont
  {Radovic}}, \bibinfo {author} {\bibfnamefont {A.}~\bibnamefont {Bostwick}},
  \bibinfo {author} {\bibfnamefont {E.}~\bibnamefont {Rotenberg}}, \bibinfo
  {author} {\bibfnamefont {A.}~\bibnamefont {Fleszar}}, \bibinfo {author}
  {\bibfnamefont {G.}~\bibnamefont {Bihlmayer}}, \bibinfo {author}
  {\bibfnamefont {D.}~\bibnamefont {Wortmann}}, \bibinfo {author}
  {\bibfnamefont {G.}~\bibnamefont {Li}}, \bibinfo {author} {\bibfnamefont
  {W.}~\bibnamefont {Hanke}}, \bibinfo {author} {\bibfnamefont
  {R.}~\bibnamefont {Claessen}}, \ and\ \bibinfo {author} {\bibfnamefont
  {J.}~\bibnamefont {Sch\"afer}},\ }\bibfield  {title} {\enquote {\bibinfo
  {title} {Elemental topological insulator with tunable fermi level: Strained
  $\ensuremath{\alpha}$-{Sn} on {InSb}(001)},}\ }\href {\doibase
  10.1103/PhysRevLett.111.157205} {\bibfield  {journal} {\bibinfo  {journal}
  {Phys. Rev. Lett.}\ }\textbf {\bibinfo {volume} {111}},\ \bibinfo {pages}
  {157205} (\bibinfo {year} {2013})}\BibitemShut {NoStop}%
\bibitem [{\citenamefont {Ohtsubo}\ \emph {et~al.}(2013)\citenamefont
  {Ohtsubo}, \citenamefont {Le~F\`evre}, \citenamefont {Bertran},\ and\
  \citenamefont {Taleb-Ibrahimi}}]{Ohtsubo2013-alphaSnARPES}%
  \BibitemOpen
  \bibfield  {author} {\bibinfo {author} {\bibfnamefont {Y.}~\bibnamefont
  {Ohtsubo}}, \bibinfo {author} {\bibfnamefont {P.}~\bibnamefont {Le~F\`evre}},
  \bibinfo {author} {\bibfnamefont {F.~m.~c.}\ \bibnamefont {Bertran}}, \ and\
  \bibinfo {author} {\bibfnamefont {A.}~\bibnamefont {Taleb-Ibrahimi}},\
  }\bibfield  {title} {\enquote {\bibinfo {title} {Dirac cone with helical spin
  polarization in ultrathin $\ensuremath{\alpha}$-{Sn}(001) films},}\ }\href
  {\doibase 10.1103/PhysRevLett.111.216401} {\bibfield  {journal} {\bibinfo
  {journal} {Phys. Rev. Lett.}\ }\textbf {\bibinfo {volume} {111}},\ \bibinfo
  {pages} {216401} (\bibinfo {year} {2013})}\BibitemShut {NoStop}%
\bibitem [{\citenamefont {Xu}\ \emph {et~al.}(2017)\citenamefont {Xu},
  \citenamefont {Chan}, \citenamefont {Chen}, \citenamefont {Chen},
  \citenamefont {Wang}, \citenamefont {Dejoie}, \citenamefont {Wong},
  \citenamefont {Hlevyack}, \citenamefont {Ryu}, \citenamefont {Kee},
  \citenamefont {Tamura}, \citenamefont {Chou}, \citenamefont {Hussain},
  \citenamefont {Mo},\ and\ \citenamefont {Chiang}}]{Xu2017-alphaSn111-semi}%
  \BibitemOpen
  \bibfield  {author} {\bibinfo {author} {\bibfnamefont {C.-Z.}\ \bibnamefont
  {Xu}}, \bibinfo {author} {\bibfnamefont {Y.-H.}\ \bibnamefont {Chan}},
  \bibinfo {author} {\bibfnamefont {Y.}~\bibnamefont {Chen}}, \bibinfo {author}
  {\bibfnamefont {P.}~\bibnamefont {Chen}}, \bibinfo {author} {\bibfnamefont
  {X.}~\bibnamefont {Wang}}, \bibinfo {author} {\bibfnamefont {C.}~\bibnamefont
  {Dejoie}}, \bibinfo {author} {\bibfnamefont {M.-H.}\ \bibnamefont {Wong}},
  \bibinfo {author} {\bibfnamefont {J.~A.}\ \bibnamefont {Hlevyack}}, \bibinfo
  {author} {\bibfnamefont {H.}~\bibnamefont {Ryu}}, \bibinfo {author}
  {\bibfnamefont {H.-Y.}\ \bibnamefont {Kee}}, \bibinfo {author} {\bibfnamefont
  {N.}~\bibnamefont {Tamura}}, \bibinfo {author} {\bibfnamefont {M.-Y.}\
  \bibnamefont {Chou}}, \bibinfo {author} {\bibfnamefont {Z.}~\bibnamefont
  {Hussain}}, \bibinfo {author} {\bibfnamefont {S.-K.}\ \bibnamefont {Mo}}, \
  and\ \bibinfo {author} {\bibfnamefont {T.-C.}\ \bibnamefont {Chiang}},\
  }\bibfield  {title} {\enquote {\bibinfo {title} {Elemental topological dirac
  semimetal: $\ensuremath{\alpha}$-{Sn} on {InSb}(111)},}\ }\href {\doibase
  10.1103/PhysRevLett.118.146402} {\bibfield  {journal} {\bibinfo  {journal}
  {Phys. Rev. Lett.}\ }\textbf {\bibinfo {volume} {118}},\ \bibinfo {pages}
  {146402} (\bibinfo {year} {2017})}\BibitemShut {NoStop}%
\bibitem [{\citenamefont {Rogalev}\ \emph {et~al.}(2017)\citenamefont
  {Rogalev}, \citenamefont {Rauch}, \citenamefont {Scholz}, \citenamefont
  {Reis}, \citenamefont {Dudy}, \citenamefont {Fleszar}, \citenamefont
  {Husanu}, \citenamefont {Strocov}, \citenamefont {Henk}, \citenamefont
  {Mertig}, \citenamefont {Sch\"afer},\ and\ \citenamefont
  {Claessen}}]{RogalevPhysRevB.95.161117}%
  \BibitemOpen
  \bibfield  {author} {\bibinfo {author} {\bibfnamefont {V.~A.}\ \bibnamefont
  {Rogalev}}, \bibinfo {author} {\bibfnamefont {T.}~\bibnamefont {Rauch}},
  \bibinfo {author} {\bibfnamefont {M.~R.}\ \bibnamefont {Scholz}}, \bibinfo
  {author} {\bibfnamefont {F.}~\bibnamefont {Reis}}, \bibinfo {author}
  {\bibfnamefont {L.}~\bibnamefont {Dudy}}, \bibinfo {author} {\bibfnamefont
  {A.}~\bibnamefont {Fleszar}}, \bibinfo {author} {\bibfnamefont {M.-A.}\
  \bibnamefont {Husanu}}, \bibinfo {author} {\bibfnamefont {V.~N.}\
  \bibnamefont {Strocov}}, \bibinfo {author} {\bibfnamefont {J.}~\bibnamefont
  {Henk}}, \bibinfo {author} {\bibfnamefont {I.}~\bibnamefont {Mertig}},
  \bibinfo {author} {\bibfnamefont {J.}~\bibnamefont {Sch\"afer}}, \ and\
  \bibinfo {author} {\bibfnamefont {R.}~\bibnamefont {Claessen}},\ }\bibfield
  {title} {\enquote {\bibinfo {title} {Double band inversion in
  $\ensuremath{\alpha}$-{Sn}: Appearance of topological surface states and the
  role of orbital composition},}\ }\href {\doibase 10.1103/PhysRevB.95.161117}
  {\bibfield  {journal} {\bibinfo  {journal} {Phys. Rev. B}\ }\textbf {\bibinfo
  {volume} {95}},\ \bibinfo {pages} {161117} (\bibinfo {year}
  {2017})}\BibitemShut {NoStop}%
\bibitem [{\citenamefont {Scholz}\ \emph {et~al.}(2018)\citenamefont {Scholz},
  \citenamefont {Rogalev}, \citenamefont {Dudy}, \citenamefont {Reis},
  \citenamefont {Adler}, \citenamefont {Aulbach}, \citenamefont
  {Collins-McIntyre}, \citenamefont {Duffy}, \citenamefont {Yang},
  \citenamefont {Chen}, \citenamefont {Hesjedal}, \citenamefont {Liu},
  \citenamefont {Hoesch}, \citenamefont {Muff}, \citenamefont {Dil},
  \citenamefont {Sch\"afer},\ and\ \citenamefont
  {Claessen}}]{Scholz-PhysRevB.97.075101}%
  \BibitemOpen
  \bibfield  {author} {\bibinfo {author} {\bibfnamefont {M.~R.}\ \bibnamefont
  {Scholz}}, \bibinfo {author} {\bibfnamefont {V.~A.}\ \bibnamefont {Rogalev}},
  \bibinfo {author} {\bibfnamefont {L.}~\bibnamefont {Dudy}}, \bibinfo {author}
  {\bibfnamefont {F.}~\bibnamefont {Reis}}, \bibinfo {author} {\bibfnamefont
  {F.}~\bibnamefont {Adler}}, \bibinfo {author} {\bibfnamefont
  {J.}~\bibnamefont {Aulbach}}, \bibinfo {author} {\bibfnamefont {L.~J.}\
  \bibnamefont {Collins-McIntyre}}, \bibinfo {author} {\bibfnamefont {L.~B.}\
  \bibnamefont {Duffy}}, \bibinfo {author} {\bibfnamefont {H.~F.}\ \bibnamefont
  {Yang}}, \bibinfo {author} {\bibfnamefont {Y.~L.}\ \bibnamefont {Chen}},
  \bibinfo {author} {\bibfnamefont {T.}~\bibnamefont {Hesjedal}}, \bibinfo
  {author} {\bibfnamefont {Z.~K.}\ \bibnamefont {Liu}}, \bibinfo {author}
  {\bibfnamefont {M.}~\bibnamefont {Hoesch}}, \bibinfo {author} {\bibfnamefont
  {S.}~\bibnamefont {Muff}}, \bibinfo {author} {\bibfnamefont {J.~H.}\
  \bibnamefont {Dil}}, \bibinfo {author} {\bibfnamefont {J.}~\bibnamefont
  {Sch\"afer}}, \ and\ \bibinfo {author} {\bibfnamefont {R.}~\bibnamefont
  {Claessen}},\ }\bibfield  {title} {\enquote {\bibinfo {title} {Topological
  surface state of $\ensuremath{\alpha}$-{Sn} on {InSb}(001) as studied by
  photoemission},}\ }\href {\doibase 10.1103/PhysRevB.97.075101} {\bibfield
  {journal} {\bibinfo  {journal} {Phys. Rev. B}\ }\textbf {\bibinfo {volume}
  {97}},\ \bibinfo {pages} {075101} (\bibinfo {year} {2018})}\BibitemShut
  {NoStop}%
\bibitem [{\citenamefont {Huang}\ and\ \citenamefont
  {Liu}(2017)}]{Huang2017_tensileZZ}%
  \BibitemOpen
  \bibfield  {author} {\bibinfo {author} {\bibfnamefont {H.}~\bibnamefont
  {Huang}}\ and\ \bibinfo {author} {\bibfnamefont {F.}~\bibnamefont {Liu}},\
  }\bibfield  {title} {\enquote {\bibinfo {title} {Tensile strained gray tin:
  Dirac semimetal for observing negative magnetoresistance with {Shubnikov}--de
  {Haas} oscillations},}\ }\href {\doibase 10.1103/PhysRevB.95.201101}
  {\bibfield  {journal} {\bibinfo  {journal} {Phys. Rev. B}\ }\textbf {\bibinfo
  {volume} {95}},\ \bibinfo {pages} {201101} (\bibinfo {year}
  {2017})}\BibitemShut {NoStop}%
\bibitem [{\citenamefont {Zhang}\ \emph {et~al.}(2018)\citenamefont {Zhang},
  \citenamefont {Wang}, \citenamefont {Ruan}, \citenamefont {Yao},\ and\
  \citenamefont {Zhang}}]{Zhang_Luttinger_Sn2018}%
  \BibitemOpen
  \bibfield  {author} {\bibinfo {author} {\bibfnamefont {D.}~\bibnamefont
  {Zhang}}, \bibinfo {author} {\bibfnamefont {H.}~\bibnamefont {Wang}},
  \bibinfo {author} {\bibfnamefont {J.}~\bibnamefont {Ruan}}, \bibinfo {author}
  {\bibfnamefont {G.}~\bibnamefont {Yao}}, \ and\ \bibinfo {author}
  {\bibfnamefont {H.}~\bibnamefont {Zhang}},\ }\bibfield  {title} {\enquote
  {\bibinfo {title} {Engineering topological phases in the {Luttinger}
  semimetal $\ensuremath{\alpha}$-{Sn}},}\ }\href {\doibase
  10.1103/PhysRevB.97.195139} {\bibfield  {journal} {\bibinfo  {journal} {Phys.
  Rev. B}\ }\textbf {\bibinfo {volume} {97}},\ \bibinfo {pages} {195139}
  (\bibinfo {year} {2018})}\BibitemShut {NoStop}%
\bibitem [{\citenamefont {Rogalev}\ \emph {et~al.}(2019)\citenamefont
  {Rogalev}, \citenamefont {Reis}, \citenamefont {Adler}, \citenamefont
  {Bauernfeind}, \citenamefont {Erhardt}, \citenamefont {Kowalewski},
  \citenamefont {Scholz}, \citenamefont {Dudy}, \citenamefont {Duffy},
  \citenamefont {Hesjedal}, \citenamefont {Hoesch}, \citenamefont {Bihlmayer},
  \citenamefont {Sch\"afer},\ and\ \citenamefont {Claessen}}]{Rogalev2019}%
  \BibitemOpen
  \bibfield  {author} {\bibinfo {author} {\bibfnamefont {V.~A.}\ \bibnamefont
  {Rogalev}}, \bibinfo {author} {\bibfnamefont {F.}~\bibnamefont {Reis}},
  \bibinfo {author} {\bibfnamefont {F.}~\bibnamefont {Adler}}, \bibinfo
  {author} {\bibfnamefont {M.}~\bibnamefont {Bauernfeind}}, \bibinfo {author}
  {\bibfnamefont {J.}~\bibnamefont {Erhardt}}, \bibinfo {author} {\bibfnamefont
  {A.}~\bibnamefont {Kowalewski}}, \bibinfo {author} {\bibfnamefont {M.~R.}\
  \bibnamefont {Scholz}}, \bibinfo {author} {\bibfnamefont {L.}~\bibnamefont
  {Dudy}}, \bibinfo {author} {\bibfnamefont {L.~B.}\ \bibnamefont {Duffy}},
  \bibinfo {author} {\bibfnamefont {T.}~\bibnamefont {Hesjedal}}, \bibinfo
  {author} {\bibfnamefont {M.}~\bibnamefont {Hoesch}}, \bibinfo {author}
  {\bibfnamefont {G.}~\bibnamefont {Bihlmayer}}, \bibinfo {author}
  {\bibfnamefont {J.}~\bibnamefont {Sch\"afer}}, \ and\ \bibinfo {author}
  {\bibfnamefont {R.}~\bibnamefont {Claessen}},\ }\bibfield  {title} {\enquote
  {\bibinfo {title} {Tailoring the topological surface state in ultrathin
  $\ensuremath{\alpha}$-{Sn}(111) films},}\ }\href {\doibase
  10.1103/PhysRevB.100.245144} {\bibfield  {journal} {\bibinfo  {journal}
  {Phys. Rev. B}\ }\textbf {\bibinfo {volume} {100}},\ \bibinfo {pages}
  {245144} (\bibinfo {year} {2019})}\BibitemShut {NoStop}%
\bibitem [{\citenamefont {Barbedienne}\ \emph {et~al.}(2018)\citenamefont
  {Barbedienne}, \citenamefont {Varignon}, \citenamefont {Reyren},
  \citenamefont {Marty}, \citenamefont {Vergnaud}, \citenamefont {Jamet},
  \citenamefont {Gomez-Carbonell}, \citenamefont {Lema\^{\i}tre}, \citenamefont
  {Le~F\`evre}, \citenamefont {Bertran}, \citenamefont {Taleb-Ibrahimi},
  \citenamefont {Jaffr\`es}, \citenamefont {George},\ and\ \citenamefont
  {Fert}}]{Barbedienne-PhysRevB.98.195445}%
  \BibitemOpen
  \bibfield  {author} {\bibinfo {author} {\bibfnamefont {Q.}~\bibnamefont
  {Barbedienne}}, \bibinfo {author} {\bibfnamefont {J.}~\bibnamefont
  {Varignon}}, \bibinfo {author} {\bibfnamefont {N.}~\bibnamefont {Reyren}},
  \bibinfo {author} {\bibfnamefont {A.}~\bibnamefont {Marty}}, \bibinfo
  {author} {\bibfnamefont {C.}~\bibnamefont {Vergnaud}}, \bibinfo {author}
  {\bibfnamefont {M.}~\bibnamefont {Jamet}}, \bibinfo {author} {\bibfnamefont
  {C.}~\bibnamefont {Gomez-Carbonell}}, \bibinfo {author} {\bibfnamefont
  {A.}~\bibnamefont {Lema\^{\i}tre}}, \bibinfo {author} {\bibfnamefont
  {P.}~\bibnamefont {Le~F\`evre}}, \bibinfo {author} {\bibfnamefont {F.~m.~c.}\
  \bibnamefont {Bertran}}, \bibinfo {author} {\bibfnamefont {A.}~\bibnamefont
  {Taleb-Ibrahimi}}, \bibinfo {author} {\bibfnamefont {H.}~\bibnamefont
  {Jaffr\`es}}, \bibinfo {author} {\bibfnamefont {J.-M.}\ \bibnamefont
  {George}}, \ and\ \bibinfo {author} {\bibfnamefont {A.}~\bibnamefont
  {Fert}},\ }\bibfield  {title} {\enquote {\bibinfo {title} {Angular-resolved
  photoemission electron spectroscopy and transport studies of the elemental
  topological insulator $\ensuremath{\alpha}$-{Sn}},}\ }\href {\doibase
  10.1103/PhysRevB.98.195445} {\bibfield  {journal} {\bibinfo  {journal} {Phys.
  Rev. B}\ }\textbf {\bibinfo {volume} {98}},\ \bibinfo {pages} {195445}
  (\bibinfo {year} {2018})}\BibitemShut {NoStop}%
\bibitem [{\citenamefont {Frost}, \citenamefont {Schindler},\ and\
  \citenamefont {Bigl}(1998)}]{Frost_1998-InSb_sput}%
  \BibitemOpen
  \bibfield  {author} {\bibinfo {author} {\bibfnamefont {F.}~\bibnamefont
  {Frost}}, \bibinfo {author} {\bibfnamefont {A.}~\bibnamefont {Schindler}}, \
  and\ \bibinfo {author} {\bibfnamefont {F.}~\bibnamefont {Bigl}},\ }\bibfield
  {title} {\enquote {\bibinfo {title} {Reactive ion beam etching of {InSb} and
  {InAs} with ultrasmooth surfaces},}\ }\href {\doibase
  10.1088/0268-1242/13/5/014} {\bibfield  {journal} {\bibinfo  {journal}
  {Semiconductor Science and Technology}\ }\textbf {\bibinfo {volume} {13}},\
  \bibinfo {pages} {523--527} (\bibinfo {year} {1998})}\BibitemShut {NoStop}%
\bibitem [{\citenamefont {Magnano}\ \emph {et~al.}(2002)\citenamefont
  {Magnano}, \citenamefont {Cepek}, \citenamefont {Gardonio}, \citenamefont
  {Allieri}, \citenamefont {Baek}, \citenamefont {Vescovo}, \citenamefont
  {Roca}, \citenamefont {Avila}, \citenamefont {Betti}, \citenamefont
  {Mariani},\ and\ \citenamefont {Sancrotti}}]{MAGNANO200229}%
  \BibitemOpen
  \bibfield  {author} {\bibinfo {author} {\bibfnamefont {E.}~\bibnamefont
  {Magnano}}, \bibinfo {author} {\bibfnamefont {C.}~\bibnamefont {Cepek}},
  \bibinfo {author} {\bibfnamefont {S.}~\bibnamefont {Gardonio}}, \bibinfo
  {author} {\bibfnamefont {B.}~\bibnamefont {Allieri}}, \bibinfo {author}
  {\bibfnamefont {I.}~\bibnamefont {Baek}}, \bibinfo {author} {\bibfnamefont
  {E.}~\bibnamefont {Vescovo}}, \bibinfo {author} {\bibfnamefont
  {L.}~\bibnamefont {Roca}}, \bibinfo {author} {\bibfnamefont {J.}~\bibnamefont
  {Avila}}, \bibinfo {author} {\bibfnamefont {M.~G.}\ \bibnamefont {Betti}},
  \bibinfo {author} {\bibfnamefont {C.}~\bibnamefont {Mariani}}, \ and\
  \bibinfo {author} {\bibfnamefont {M.}~\bibnamefont {Sancrotti}},\ }\bibfield
  {title} {\enquote {\bibinfo {title} {Sn on {InSb}(100)–c($2\times8$):
  growth morphology and electronic structure},}\ }\href {\doibase
  https://doi.org/10.1016/S0368-2048(02)00169-X} {\bibfield  {journal}
  {\bibinfo  {journal} {J Electron Spectrosc}\ }\textbf {\bibinfo {volume}
  {127}},\ \bibinfo {pages} {29} (\bibinfo {year} {2002})}\BibitemShut
  {NoStop}%
\bibitem [{\citenamefont {Kahn}(1983)}]{SemicondSurfaces-KAHN1983193}%
  \BibitemOpen
  \bibfield  {author} {\bibinfo {author} {\bibfnamefont {A.}~\bibnamefont
  {Kahn}},\ }\bibfield  {title} {\enquote {\bibinfo {title} {Semiconductor
  surface structures},}\ }\href {\doibase
  https://doi.org/10.1016/0167-5729(83)90006-7} {\bibfield  {journal} {\bibinfo
   {journal} {Surface Science Reports}\ }\textbf {\bibinfo {volume} {3}},\
  \bibinfo {pages} {193--300} (\bibinfo {year} {1983})}\BibitemShut {NoStop}%
\bibitem [{\citenamefont {Tromp}, \citenamefont {Hamers},\ and\ \citenamefont
  {Demuth}(1985)}]{SiDimer1985PhysRevLett.55.1303}%
  \BibitemOpen
  \bibfield  {author} {\bibinfo {author} {\bibfnamefont {R.~M.}\ \bibnamefont
  {Tromp}}, \bibinfo {author} {\bibfnamefont {R.~J.}\ \bibnamefont {Hamers}}, \
  and\ \bibinfo {author} {\bibfnamefont {J.~E.}\ \bibnamefont {Demuth}},\
  }\bibfield  {title} {\enquote {\bibinfo {title} {Si(001) dimer structure
  observed with scanning tunneling microscopy},}\ }\href {\doibase
  10.1103/PhysRevLett.55.1303} {\bibfield  {journal} {\bibinfo  {journal}
  {Phys. Rev. Lett.}\ }\textbf {\bibinfo {volume} {55}},\ \bibinfo {pages}
  {1303--1306} (\bibinfo {year} {1985})}\BibitemShut {NoStop}%
\bibitem [{\citenamefont {Ramstad}, \citenamefont {Brocks},\ and\ \citenamefont
  {Kelly}(1995)}]{SiDimerTheoPhysRevB.51.14504}%
  \BibitemOpen
  \bibfield  {author} {\bibinfo {author} {\bibfnamefont {A.}~\bibnamefont
  {Ramstad}}, \bibinfo {author} {\bibfnamefont {G.}~\bibnamefont {Brocks}}, \
  and\ \bibinfo {author} {\bibfnamefont {P.~J.}\ \bibnamefont {Kelly}},\
  }\bibfield  {title} {\enquote {\bibinfo {title} {Theoretical study of the
  {Si}(100) surface reconstruction},}\ }\href {\doibase
  10.1103/PhysRevB.51.14504} {\bibfield  {journal} {\bibinfo  {journal} {Phys.
  Rev. B}\ }\textbf {\bibinfo {volume} {51}},\ \bibinfo {pages} {14504--14523}
  (\bibinfo {year} {1995})}\BibitemShut {NoStop}%
\bibitem [{\citenamefont {Kubby}\ and\ \citenamefont
  {Boland}(1996)}]{STMsemi-KUBBY199661}%
  \BibitemOpen
  \bibfield  {author} {\bibinfo {author} {\bibfnamefont {J.}~\bibnamefont
  {Kubby}}\ and\ \bibinfo {author} {\bibfnamefont {J.}~\bibnamefont {Boland}},\
  }\bibfield  {title} {\enquote {\bibinfo {title} {Scanning tunneling
  microscopy of semiconductor surfaces},}\ }\href {\doibase
  https://doi.org/10.1016/S0167-5729(97)80001-5} {\bibfield  {journal}
  {\bibinfo  {journal} {Surface Science Reports}\ }\textbf {\bibinfo {volume}
  {26}},\ \bibinfo {pages} {61--204} (\bibinfo {year} {1996})}\BibitemShut
  {NoStop}%
\bibitem [{\citenamefont {Berchenko}\ \emph {et~al.}(2018)\citenamefont
  {Berchenko}, \citenamefont {Vitchev}, \citenamefont {Trzyna}, \citenamefont
  {Wojnarowska-Nowak}, \citenamefont {Szczerbakow}, \citenamefont {Badyla},
  \citenamefont {Cebulski},\ and\ \citenamefont
  {Story}}]{SnTeOXY_BERCHENKO2018134}%
  \BibitemOpen
  \bibfield  {author} {\bibinfo {author} {\bibfnamefont {N.}~\bibnamefont
  {Berchenko}}, \bibinfo {author} {\bibfnamefont {R.}~\bibnamefont {Vitchev}},
  \bibinfo {author} {\bibfnamefont {M.}~\bibnamefont {Trzyna}}, \bibinfo
  {author} {\bibfnamefont {R.}~\bibnamefont {Wojnarowska-Nowak}}, \bibinfo
  {author} {\bibfnamefont {A.}~\bibnamefont {Szczerbakow}}, \bibinfo {author}
  {\bibfnamefont {A.}~\bibnamefont {Badyla}}, \bibinfo {author} {\bibfnamefont
  {J.}~\bibnamefont {Cebulski}}, \ and\ \bibinfo {author} {\bibfnamefont
  {T.}~\bibnamefont {Story}},\ }\bibfield  {title} {\enquote {\bibinfo {title}
  {Surface oxidation of {SnTe} topological crystalline insulator},}\ }\href
  {\doibase https://doi.org/10.1016/j.apsusc.2018.04.246} {\bibfield  {journal}
  {\bibinfo  {journal} {Applied Surface Science}\ }\textbf {\bibinfo {volume}
  {452}},\ \bibinfo {pages} {134--140} (\bibinfo {year} {2018})}\BibitemShut
  {NoStop}%
\bibitem [{\citenamefont {Liu}\ \emph {et~al.}(2019)\citenamefont {Liu},
  \citenamefont {Xie}, \citenamefont {Miller}, \citenamefont {Ebine},
  \citenamefont {Kumaravadivel}, \citenamefont {Sohn},\ and\ \citenamefont
  {Cha}}]{SnTe_degrad_LIU2019351}%
  \BibitemOpen
  \bibfield  {author} {\bibinfo {author} {\bibfnamefont {P.}~\bibnamefont
  {Liu}}, \bibinfo {author} {\bibfnamefont {Y.}~\bibnamefont {Xie}}, \bibinfo
  {author} {\bibfnamefont {E.}~\bibnamefont {Miller}}, \bibinfo {author}
  {\bibfnamefont {Y.}~\bibnamefont {Ebine}}, \bibinfo {author} {\bibfnamefont
  {P.}~\bibnamefont {Kumaravadivel}}, \bibinfo {author} {\bibfnamefont
  {S.}~\bibnamefont {Sohn}}, \ and\ \bibinfo {author} {\bibfnamefont {J.~J.}\
  \bibnamefont {Cha}},\ }\bibfield  {title} {\enquote {\bibinfo {title}
  {Dislocation-driven {SnTe} surface defects during chemical vapor deposition
  growth},}\ }\href {\doibase https://doi.org/10.1016/j.jpcs.2017.12.016}
  {\bibfield  {journal} {\bibinfo  {journal} {Journal of Physics and Chemistry
  of Solids}\ }\textbf {\bibinfo {volume} {128}},\ \bibinfo {pages} {351--359}
  (\bibinfo {year} {2019})}\BibitemShut {NoStop}%
\bibitem [{\citenamefont {Valassiades}\ and\ \citenamefont
  {Economou}(1975)}]{SnTeMOSS_FANO1977467}%
  \BibitemOpen
  \bibfield  {author} {\bibinfo {author} {\bibfnamefont {O.}~\bibnamefont
  {Valassiades}}\ and\ \bibinfo {author} {\bibfnamefont {N.~A.}\ \bibnamefont
  {Economou}},\ }\bibfield  {title} {\enquote {\bibinfo {title} {On the phase
  transformation of {SnTe}},}\ }\href {\doibase 10.1002/pssa.2210300119}
  {\bibfield  {journal} {\bibinfo  {journal} {physica status solidi (a)}\
  }\textbf {\bibinfo {volume} {30}},\ \bibinfo {pages} {187--195} (\bibinfo
  {year} {1975})}\BibitemShut {NoStop}%
\bibitem [{SnT()}]{SnTe_Landolt1998}%
  \BibitemOpen
  \bibfield  {title} {\enquote {\bibinfo {title} {Tin telluride ({SnTe})
  crystal structure, lattice parameters},}\ }in\ \href {\doibase
  10.1007/10681727_862} {\emph {\bibinfo {booktitle} {Non-Tetrahedrally Bonded
  Elements and Binary Compounds I}}},\ \bibinfo {editor} {edited by\ \bibinfo
  {editor} {\bibfnamefont {O.}~\bibnamefont {Madelung}}, \bibinfo {editor}
  {\bibfnamefont {U.}~\bibnamefont {R{\"o}ssler}}, \ and\ \bibinfo {editor}
  {\bibfnamefont {M.}~\bibnamefont {Schulz}}}\ (\bibinfo  {publisher}
  {Springer-Verlag Berlin Heidelberg})\ pp.\ \bibinfo {pages} {1--8},\ \bibinfo
  {note} {copyright 1998 Springer-Verlag Berlin Heidelberg}\BibitemShut
  {NoStop}%
\bibitem [{\citenamefont {Li}(2020)}]{li2020chiral}%
  \BibitemOpen
  \bibfield  {author} {\bibinfo {author} {\bibfnamefont {Q.}~\bibnamefont
  {Li}},\ }\bibfield  {title} {\enquote {\bibinfo {title} {The chiral qubit:
  quantum computing with chiral anomaly},}\ }\href@noop {} {\bibfield
  {journal} {\bibinfo  {journal} {Bulletin of the American Physical Society}\ }
  (\bibinfo {year} {2020})}\BibitemShut {NoStop}%
\bibitem [{\citenamefont {Philip}, \citenamefont {Kaushik},\ and\ \citenamefont
  {Kharzeev}(2020)}]{philip2020chiral}%
  \BibitemOpen
  \bibfield  {author} {\bibinfo {author} {\bibfnamefont {E.}~\bibnamefont
  {Philip}}, \bibinfo {author} {\bibfnamefont {S.}~\bibnamefont {Kaushik}}, \
  and\ \bibinfo {author} {\bibfnamefont {D.}~\bibnamefont {Kharzeev}},\
  }\bibfield  {title} {\enquote {\bibinfo {title} {Chiral qubit: implimenting a
  qubit using chiral charge and chiral anomaly},}\ }\href@noop {} {\bibfield
  {journal} {\bibinfo  {journal} {Bulletin of the American Physical Society}\ }
  (\bibinfo {year} {2020})}\BibitemShut {NoStop}%
\bibitem [{\citenamefont {Greenwood}\ and\ \citenamefont
  {Gibb}(1971)}]{Greenwood1971}%
  \BibitemOpen
  \bibfield  {author} {\bibinfo {author} {\bibfnamefont {N.~N.}\ \bibnamefont
  {Greenwood}}\ and\ \bibinfo {author} {\bibfnamefont {T.~C.}\ \bibnamefont
  {Gibb}},\ }\href {\doibase 10.1007/978-94-009-5697-1} {\emph {\bibinfo
  {title} {M\"{o}ssbauer Spectroscopy}}}\ (\bibinfo  {publisher} {Springer
  Netherlands},\ \bibinfo {year} {1971})\BibitemShut {NoStop}%
\bibitem [{\citenamefont {Stevens}(1983)}]{Stevens1983}%
  \BibitemOpen
  \bibfield  {author} {\bibinfo {author} {\bibfnamefont {J.~G.}\ \bibnamefont
  {Stevens}},\ }\bibfield  {title} {\enquote {\bibinfo {title} {Isomer shift
  reference scales},}\ }\href {\doibase 10.1007/BF01027252} {\bibfield
  {journal} {\bibinfo  {journal} {Hyperfine Interactions}\ }\textbf {\bibinfo
  {volume} {13}},\ \bibinfo {pages} {221--236} (\bibinfo {year}
  {1983})}\BibitemShut {NoStop}%
\bibitem [{\citenamefont {Svane}\ \emph {et~al.}(1997)\citenamefont {Svane},
  \citenamefont {Christensen}, \citenamefont {Rodriguez},\ and\ \citenamefont
  {Methfessel}}]{Svane1997}%
  \BibitemOpen
  \bibfield  {author} {\bibinfo {author} {\bibfnamefont {A.}~\bibnamefont
  {Svane}}, \bibinfo {author} {\bibfnamefont {N.~E.}\ \bibnamefont
  {Christensen}}, \bibinfo {author} {\bibfnamefont {C.~O.}\ \bibnamefont
  {Rodriguez}}, \ and\ \bibinfo {author} {\bibfnamefont {M.}~\bibnamefont
  {Methfessel}},\ }\bibfield  {title} {\enquote {\bibinfo {title} {Calculations
  of hyperfine parameters in tin compounds},}\ }\href {\doibase
  10.1103/physrevb.55.12572} {\bibfield  {journal} {\bibinfo  {journal}
  {Physical Review B}\ }\textbf {\bibinfo {volume} {55}},\ \bibinfo {pages}
  {12572--12577} (\bibinfo {year} {1997})}\BibitemShut {NoStop}%
\bibitem [{\citenamefont {Skwarek}\ \emph {et~al.}(2016)\citenamefont
  {Skwarek}, \citenamefont {Zachariasz}, \citenamefont {Żukrowski},
  \citenamefont {Synkiewicz},\ and\ \citenamefont
  {Witek}}]{Sn-bulk_SKWAREK201610}%
  \BibitemOpen
  \bibfield  {author} {\bibinfo {author} {\bibfnamefont {A.}~\bibnamefont
  {Skwarek}}, \bibinfo {author} {\bibfnamefont {P.}~\bibnamefont {Zachariasz}},
  \bibinfo {author} {\bibfnamefont {J.}~\bibnamefont {Żukrowski}}, \bibinfo
  {author} {\bibfnamefont {B.}~\bibnamefont {Synkiewicz}}, \ and\ \bibinfo
  {author} {\bibfnamefont {K.}~\bibnamefont {Witek}},\ }\bibfield  {title}
  {\enquote {\bibinfo {title} {Early stage detection of \mbox{$\beta
  \rightarrow \alpha$} transition in {Sn} by {Mössbauer} spectroscopy},}\
  }\href {\doibase https://doi.org/10.1016/j.matchemphys.2016.07.061}
  {\bibfield  {journal} {\bibinfo  {journal} {Materials Chemistry and Physics}\
  }\textbf {\bibinfo {volume} {182}},\ \bibinfo {pages} {10--14} (\bibinfo
  {year} {2016})}\BibitemShut {NoStop}%
\bibitem [{\citenamefont {Zachariasz}\ \emph {et~al.}(2018)\citenamefont
  {Zachariasz}, \citenamefont {Skwarek}, \citenamefont {Illés}, \citenamefont
  {Żukrowski}, \citenamefont {Hurtony},\ and\ \citenamefont
  {Witek}}]{Sn-bulk_ZACHARIASZ2018165}%
  \BibitemOpen
  \bibfield  {author} {\bibinfo {author} {\bibfnamefont {P.}~\bibnamefont
  {Zachariasz}}, \bibinfo {author} {\bibfnamefont {A.}~\bibnamefont {Skwarek}},
  \bibinfo {author} {\bibfnamefont {B.}~\bibnamefont {Illés}}, \bibinfo
  {author} {\bibfnamefont {J.}~\bibnamefont {Żukrowski}}, \bibinfo {author}
  {\bibfnamefont {T.}~\bibnamefont {Hurtony}}, \ and\ \bibinfo {author}
  {\bibfnamefont {K.}~\bibnamefont {Witek}},\ }\bibfield  {title} {\enquote
  {\bibinfo {title} {Mössbauer studies of \mbox{$\beta \rightarrow \alpha$}
  phase transition in {Sn}-rich solder alloys},}\ }\href {\doibase
  https://doi.org/10.1016/j.microrel.2018.01.016} {\bibfield  {journal}
  {\bibinfo  {journal} {Microelectronics Reliability}\ }\textbf {\bibinfo
  {volume} {82}},\ \bibinfo {pages} {165--170} (\bibinfo {year}
  {2018})}\BibitemShut {NoStop}%
\bibitem [{\citenamefont {G\'omez}\ \emph {et~al.}(2003)\citenamefont
  {G\'omez}, \citenamefont {Guenzburger}, \citenamefont {Ellis}, \citenamefont
  {Hu}, \citenamefont {Alp}, \citenamefont {Baggio-Saitovitch}, \citenamefont
  {Passamani}, \citenamefont {Ketterson},\ and\ \citenamefont
  {Cho}}]{Gomez_CdTe_PhysRevB.67.115340}%
  \BibitemOpen
  \bibfield  {author} {\bibinfo {author} {\bibfnamefont {J.~A.}\ \bibnamefont
  {G\'omez}}, \bibinfo {author} {\bibfnamefont {D.}~\bibnamefont
  {Guenzburger}}, \bibinfo {author} {\bibfnamefont {D.~E.}\ \bibnamefont
  {Ellis}}, \bibinfo {author} {\bibfnamefont {M.~Y.}\ \bibnamefont {Hu}},
  \bibinfo {author} {\bibfnamefont {E.}~\bibnamefont {Alp}}, \bibinfo {author}
  {\bibfnamefont {E.~M.}\ \bibnamefont {Baggio-Saitovitch}}, \bibinfo {author}
  {\bibfnamefont {E.~C.}\ \bibnamefont {Passamani}}, \bibinfo {author}
  {\bibfnamefont {J.~B.}\ \bibnamefont {Ketterson}}, \ and\ \bibinfo {author}
  {\bibfnamefont {S.}~\bibnamefont {Cho}},\ }\bibfield  {title} {\enquote
  {\bibinfo {title} {Theoretical and experimental study of
  \ensuremath{\alpha}-{Sn} deposited on {CdTe}(001)},}\ }\href {\doibase
  10.1103/PhysRevB.67.115340} {\bibfield  {journal} {\bibinfo  {journal} {Phys.
  Rev. B}\ }\textbf {\bibinfo {volume} {67}},\ \bibinfo {pages} {115340}
  (\bibinfo {year} {2003})}\BibitemShut {NoStop}%
\bibitem [{\citenamefont {Houben}\ \emph {et~al.}(2019)\citenamefont {Houben},
  \citenamefont {Jochum}, \citenamefont {Lozano}, \citenamefont {Bisht},
  \citenamefont {Men\'endez}, \citenamefont {Merkel}, \citenamefont {R\"uffer},
  \citenamefont {Chumakov}, \citenamefont {Roelants}, \citenamefont {Partoens},
  \citenamefont {Milo\ifmmode \check{s}\else
  \v{s}\fi{}evi\ifmmode~\acute{c}\else \'{c}\fi{}}, \citenamefont {Peeters},
  \citenamefont {Couet}, \citenamefont {Vantomme}, \citenamefont {Temst},\ and\
  \citenamefont {Van~Bael}}]{Kelly-PhysRevB.100.075408}%
  \BibitemOpen
  \bibfield  {author} {\bibinfo {author} {\bibfnamefont {K.}~\bibnamefont
  {Houben}}, \bibinfo {author} {\bibfnamefont {J.~K.}\ \bibnamefont {Jochum}},
  \bibinfo {author} {\bibfnamefont {D.~P.}\ \bibnamefont {Lozano}}, \bibinfo
  {author} {\bibfnamefont {M.}~\bibnamefont {Bisht}}, \bibinfo {author}
  {\bibfnamefont {E.}~\bibnamefont {Men\'endez}}, \bibinfo {author}
  {\bibfnamefont {D.~G.}\ \bibnamefont {Merkel}}, \bibinfo {author}
  {\bibfnamefont {R.}~\bibnamefont {R\"uffer}}, \bibinfo {author}
  {\bibfnamefont {A.~I.}\ \bibnamefont {Chumakov}}, \bibinfo {author}
  {\bibfnamefont {S.}~\bibnamefont {Roelants}}, \bibinfo {author}
  {\bibfnamefont {B.}~\bibnamefont {Partoens}}, \bibinfo {author}
  {\bibfnamefont {M.~V.}\ \bibnamefont {Milo\ifmmode \check{s}\else
  \v{s}\fi{}evi\ifmmode~\acute{c}\else \'{c}\fi{}}}, \bibinfo {author}
  {\bibfnamefont {F.~M.}\ \bibnamefont {Peeters}}, \bibinfo {author}
  {\bibfnamefont {S.}~\bibnamefont {Couet}}, \bibinfo {author} {\bibfnamefont
  {A.}~\bibnamefont {Vantomme}}, \bibinfo {author} {\bibfnamefont
  {K.}~\bibnamefont {Temst}}, \ and\ \bibinfo {author} {\bibfnamefont {M.~J.}\
  \bibnamefont {Van~Bael}},\ }\bibfield  {title} {\enquote {\bibinfo {title}
  {In situ study of the $\ensuremath{\alpha}$-{Sn} to $\ensuremath{\beta}$-{Sn}
  phase transition in low-dimensional systems: Phonon behavior and
  thermodynamic properties},}\ }\href {\doibase 10.1103/PhysRevB.100.075408}
  {\bibfield  {journal} {\bibinfo  {journal} {Phys. Rev. B}\ }\textbf {\bibinfo
  {volume} {100}},\ \bibinfo {pages} {075408} (\bibinfo {year}
  {2019})}\BibitemShut {NoStop}%
\bibitem [{\citenamefont {Greenwood}, \citenamefont {Perkins},\ and\
  \citenamefont {Wall}(1967)}]{Greenwood1967symp}%
  \BibitemOpen
  \bibfield  {author} {\bibinfo {author} {\bibfnamefont {N.~N.}\ \bibnamefont
  {Greenwood}}, \bibinfo {author} {\bibfnamefont {P.~G.}\ \bibnamefont
  {Perkins}}, \ and\ \bibinfo {author} {\bibfnamefont {D.~H.}\ \bibnamefont
  {Wall}},\ }\bibfield  {title} {\enquote {\bibinfo {title} {Calculation of
  chemical shifts in the {Mössbauer} spectra of some tin({IV}) compounds},}\
  }\href {\doibase 10.1039/SF9670100051} {\bibfield  {journal} {\bibinfo
  {journal} {Symp. Faraday Soc.}\ }\textbf {\bibinfo {volume} {1}},\ \bibinfo
  {pages} {51--59} (\bibinfo {year} {1967})}\BibitemShut {NoStop}%
\bibitem [{\citenamefont {Greenwood}, \citenamefont {Perkins},\ and\
  \citenamefont {Wall}(1968)}]{GREENWOOD1968339}%
  \BibitemOpen
  \bibfield  {author} {\bibinfo {author} {\bibfnamefont {N.}~\bibnamefont
  {Greenwood}}, \bibinfo {author} {\bibfnamefont {P.}~\bibnamefont {Perkins}},
  \ and\ \bibinfo {author} {\bibfnamefont {D.}~\bibnamefont {Wall}},\
  }\bibfield  {title} {\enquote {\bibinfo {title} {The sign and magnitude of
  {$\Delta R/R$} for the 119{Sn} nucleus on excitation to the 23.8 ke{V}
  level},}\ }\href {\doibase https://doi.org/10.1016/0375-9601(68)90322-8}
  {\bibfield  {journal} {\bibinfo  {journal} {Physics Letters A}\ }\textbf
  {\bibinfo {volume} {28}},\ \bibinfo {pages} {339--340} (\bibinfo {year}
  {1968})}\BibitemShut {NoStop}%
\bibitem [{\citenamefont {Unzueta}\ \emph {et~al.}(2018)\citenamefont
  {Unzueta}, \citenamefont {L{\'{o}}pez-Garc{\'{\i}}a}, \citenamefont
  {S{\'{a}}nchez-Alarcos}, \citenamefont {Recarte}, \citenamefont
  {P{\'{e}}rez-Landaz{\'{a}}bal}, \citenamefont
  {Rodr{\'{\i}}guez-Velamaz{\'{a}}n}, \citenamefont {Garitaonandia},
  \citenamefont {Garc{\'{\i}}a},\ and\ \citenamefont {Plazaola}}]{Unzueta2018}%
  \BibitemOpen
  \bibfield  {author} {\bibinfo {author} {\bibfnamefont {I.}~\bibnamefont
  {Unzueta}}, \bibinfo {author} {\bibfnamefont {J.}~\bibnamefont
  {L{\'{o}}pez-Garc{\'{\i}}a}}, \bibinfo {author} {\bibfnamefont
  {V.}~\bibnamefont {S{\'{a}}nchez-Alarcos}}, \bibinfo {author} {\bibfnamefont
  {V.}~\bibnamefont {Recarte}}, \bibinfo {author} {\bibfnamefont {J.~I.}\
  \bibnamefont {P{\'{e}}rez-Landaz{\'{a}}bal}}, \bibinfo {author}
  {\bibfnamefont {J.~A.}\ \bibnamefont {Rodr{\'{\i}}guez-Velamaz{\'{a}}n}},
  \bibinfo {author} {\bibfnamefont {J.~S.}\ \bibnamefont {Garitaonandia}},
  \bibinfo {author} {\bibfnamefont {J.~A.}\ \bibnamefont {Garc{\'{\i}}a}}, \
  and\ \bibinfo {author} {\bibfnamefont {F.}~\bibnamefont {Plazaola}},\
  }\bibfield  {title} {\enquote {\bibinfo {title} {119{Sn} {M\"{o}ssbauer}
  spectroscopy in the study of metamagnetic shape memory alloys},}\ }\href
  {\doibase 10.1007/s10751-018-1509-z} {\bibfield  {journal} {\bibinfo
  {journal} {Hyperfine Interactions}\ }\textbf {\bibinfo {volume} {239}},\
  \bibinfo {pages} {34} (\bibinfo {year} {2018})}\BibitemShut {NoStop}%
\bibitem [{\citenamefont {Brownlee}(1950)}]{Alpha_latticeBROWNLEE1950}%
  \BibitemOpen
  \bibfield  {author} {\bibinfo {author} {\bibfnamefont {L.~D.}\ \bibnamefont
  {Brownlee}},\ }\bibfield  {title} {\enquote {\bibinfo {title} {Lattice
  constant of grey tin},}\ }\href {\doibase 10.1038/166482a0} {\bibfield
  {journal} {\bibinfo  {journal} {Nature}\ }\textbf {\bibinfo {volume} {166}},\
  \bibinfo {pages} {482--482} (\bibinfo {year} {1950})}\BibitemShut {NoStop}%
\bibitem [{\citenamefont {Thewlis}\ and\ \citenamefont
  {Davey}(1954)}]{Alpha_lattice1954}%
  \BibitemOpen
  \bibfield  {author} {\bibinfo {author} {\bibfnamefont {J.}~\bibnamefont
  {Thewlis}}\ and\ \bibinfo {author} {\bibfnamefont {A.}~\bibnamefont
  {Davey}},\ }\bibfield  {title} {\enquote {\bibinfo {title} {Thermal expansion
  of grey tin},}\ }\href@noop {} {\bibfield  {journal} {\bibinfo  {journal}
  {Nature}\ }\textbf {\bibinfo {volume} {174}},\ \bibinfo {pages} {1011--1011}
  (\bibinfo {year} {1954})}\BibitemShut {NoStop}%
\bibitem [{\citenamefont {Wyckoff}(1963)}]{wyckoff1963crystal}%
  \BibitemOpen
  \bibfield  {author} {\bibinfo {author} {\bibfnamefont {R.}~\bibnamefont
  {Wyckoff}},\ }\href {https://books.google.be/books?id=40uGpwAACAAJ} {\emph
  {\bibinfo {title} {Crystal structures}}},\ \bibinfo {series} {Crystal
  Structures}\ No.\ \bibinfo {number} {v. 1}\ (\bibinfo  {publisher}
  {Interscience Publishers},\ \bibinfo {year} {1963})\BibitemShut {NoStop}%
\bibitem [{\citenamefont {Liu}\ and\ \citenamefont
  {Peretti}(1951)}]{Liu_InSbLAT1951}%
  \BibitemOpen
  \bibfield  {author} {\bibinfo {author} {\bibfnamefont {T.~S.}\ \bibnamefont
  {Liu}}\ and\ \bibinfo {author} {\bibfnamefont {E.~A.}\ \bibnamefont
  {Peretti}},\ }\bibfield  {title} {\enquote {\bibinfo {title} {Lattice
  parameter of {InSb}},}\ }\href {\doibase 10.1007/BF03397373} {\bibfield
  {journal} {\bibinfo  {journal} {JOM}\ }\textbf {\bibinfo {volume} {3}},\
  \bibinfo {pages} {791--791} (\bibinfo {year} {1951})}\BibitemShut {NoStop}%
\bibitem [{\citenamefont {Willardson}(1968)}]{willardson1968InSbLAT}%
  \BibitemOpen
  \bibfield  {author} {\bibinfo {author} {\bibfnamefont {R.}~\bibnamefont
  {Willardson}},\ }\href@noop {} {\emph {\bibinfo {title} {Physics of III-V
  compounds}}}\ (\bibinfo  {publisher} {Academic Press},\ \bibinfo {address}
  {New York},\ \bibinfo {year} {1968})\BibitemShut {NoStop}%
\bibitem [{\citenamefont {Adachi}(1999)}]{Adachi_InSbLAT1999}%
  \BibitemOpen
  \bibfield  {author} {\bibinfo {author} {\bibfnamefont {S.}~\bibnamefont
  {Adachi}},\ }\bibfield  {title} {\enquote {\bibinfo {title} {Indium
  antimonide ({InSb})},}\ }in\ \href {\doibase 10.1007/978-1-4615-5247-5_27}
  {\emph {\bibinfo {booktitle} {Optical Constants of Crystalline and Amorphous
  Semiconductors}}}\ (\bibinfo  {publisher} {Springer {US}},\ \bibinfo {year}
  {1999})\ pp.\ \bibinfo {pages} {268--278}\BibitemShut {NoStop}%
\bibitem [{\citenamefont {Neuberger}(2013)}]{InSb_LAT_III-Vbook}%
  \BibitemOpen
  \bibfield  {author} {\bibinfo {author} {\bibfnamefont {M.}~\bibnamefont
  {Neuberger}},\ }\href@noop {} {\emph {\bibinfo {title} {III-V Semiconducting
  Compounds}}}\ (\bibinfo  {publisher} {Springer},\ \bibinfo {year}
  {2013})\BibitemShut {NoStop}%
\bibitem [{\citenamefont {Asom}\ \emph {et~al.}(1989)\citenamefont {Asom},
  \citenamefont {Kortan}, \citenamefont {Kimerling},\ and\ \citenamefont
  {Farrow}}]{AsomFarrow1989}%
  \BibitemOpen
  \bibfield  {author} {\bibinfo {author} {\bibfnamefont {M.~T.}\ \bibnamefont
  {Asom}}, \bibinfo {author} {\bibfnamefont {A.~R.}\ \bibnamefont {Kortan}},
  \bibinfo {author} {\bibfnamefont {L.~C.}\ \bibnamefont {Kimerling}}, \ and\
  \bibinfo {author} {\bibfnamefont {R.~C.}\ \bibnamefont {Farrow}},\ }\bibfield
   {title} {\enquote {\bibinfo {title} {Structure and stability of metastable
  \mbox{$\alpha$-Sn}},}\ }\href {\doibase 10.1063/1.101580} {\bibfield
  {journal} {\bibinfo  {journal} {Applied Physics Letters}\ }\textbf {\bibinfo
  {volume} {55}},\ \bibinfo {pages} {1439--1441} (\bibinfo {year}
  {1989})}\BibitemShut {NoStop}%
\bibitem [{\citenamefont {Reno}\ and\ \citenamefont
  {Stephenson}(1989)}]{vsTEMP_Reno1989}%
  \BibitemOpen
  \bibfield  {author} {\bibinfo {author} {\bibfnamefont {J.~L.}\ \bibnamefont
  {Reno}}\ and\ \bibinfo {author} {\bibfnamefont {L.~L.}\ \bibnamefont
  {Stephenson}},\ }\bibfield  {title} {\enquote {\bibinfo {title} {Effect of
  growth conditions on the stability of \mbox{$\alpha$-Sn} grown on {CdTe} by
  molecular beam epitaxy},}\ }\href {\doibase 10.1063/1.101125} {\bibfield
  {journal} {\bibinfo  {journal} {Applied Physics Letters}\ }\textbf {\bibinfo
  {volume} {54}},\ \bibinfo {pages} {2207--2209} (\bibinfo {year}
  {1989})}\BibitemShut {NoStop}%
\bibitem [{\citenamefont {Song}\ \emph {et~al.}(2019)\citenamefont {Song},
  \citenamefont {Yao}, \citenamefont {Ding}, \citenamefont {Gu}, \citenamefont
  {Deng}, \citenamefont {Lu}, \citenamefont {Lu},\ and\ \citenamefont
  {Chen}}]{vsTEMP_Song2019}%
  \BibitemOpen
  \bibfield  {author} {\bibinfo {author} {\bibfnamefont {H.}~\bibnamefont
  {Song}}, \bibinfo {author} {\bibfnamefont {J.}~\bibnamefont {Yao}}, \bibinfo
  {author} {\bibfnamefont {Y.}~\bibnamefont {Ding}}, \bibinfo {author}
  {\bibfnamefont {Y.}~\bibnamefont {Gu}}, \bibinfo {author} {\bibfnamefont
  {Y.}~\bibnamefont {Deng}}, \bibinfo {author} {\bibfnamefont {M.-H.}\
  \bibnamefont {Lu}}, \bibinfo {author} {\bibfnamefont {H.}~\bibnamefont {Lu}},
  \ and\ \bibinfo {author} {\bibfnamefont {Y.-F.}\ \bibnamefont {Chen}},\
  }\bibfield  {title} {\enquote {\bibinfo {title} {Thermal stability
  enhancement in epitaxial alpha tin films by strain engineering},}\ }\href
  {\doibase 10.1002/adem.201900410} {\bibfield  {journal} {\bibinfo  {journal}
  {Advanced Engineering Materials}\ }\textbf {\bibinfo {volume} {21}},\
  \bibinfo {pages} {1900410} (\bibinfo {year} {2019})}\BibitemShut {NoStop}%
\bibitem [{\citenamefont {Bando}\ \emph {et~al.}(2000)\citenamefont {Bando},
  \citenamefont {Koizumi}, \citenamefont {Oikawa}, \citenamefont {Daikohara},
  \citenamefont {Kulbachinskii},\ and\ \citenamefont
  {Ozaki}}]{BiTe1_Bando_2000}%
  \BibitemOpen
  \bibfield  {author} {\bibinfo {author} {\bibfnamefont {H.}~\bibnamefont
  {Bando}}, \bibinfo {author} {\bibfnamefont {K.}~\bibnamefont {Koizumi}},
  \bibinfo {author} {\bibfnamefont {Y.}~\bibnamefont {Oikawa}}, \bibinfo
  {author} {\bibfnamefont {K.}~\bibnamefont {Daikohara}}, \bibinfo {author}
  {\bibfnamefont {V.~A.}\ \bibnamefont {Kulbachinskii}}, \ and\ \bibinfo
  {author} {\bibfnamefont {H.}~\bibnamefont {Ozaki}},\ }\bibfield  {title}
  {\enquote {\bibinfo {title} {The time-dependent process of oxidation of the
  surface of {Bi2Te3} studied by x-ray photoelectron spectroscopy},}\ }\href
  {\doibase 10.1088/0953-8984/12/26/307} {\bibfield  {journal} {\bibinfo
  {journal} {Journal of Physics: Condensed Matter}\ }\textbf {\bibinfo {volume}
  {12}},\ \bibinfo {pages} {5607--5616} (\bibinfo {year} {2000})}\BibitemShut
  {NoStop}%
\bibitem [{\citenamefont {Netsou}\ \emph {et~al.}(2017)\citenamefont {Netsou},
  \citenamefont {Thupakula}, \citenamefont {Debehets}, \citenamefont {Chen},
  \citenamefont {Hirsch}, \citenamefont {Volodin}, \citenamefont {Li},
  \citenamefont {Song}, \citenamefont {Seo}, \citenamefont {Feyter},
  \citenamefont {Schouteden},\ and\ \citenamefont
  {Haesendonck}}]{BiTe2_Netsou_2017}%
  \BibitemOpen
  \bibfield  {author} {\bibinfo {author} {\bibfnamefont {A.-M.}\ \bibnamefont
  {Netsou}}, \bibinfo {author} {\bibfnamefont {U.}~\bibnamefont {Thupakula}},
  \bibinfo {author} {\bibfnamefont {J.}~\bibnamefont {Debehets}}, \bibinfo
  {author} {\bibfnamefont {T.}~\bibnamefont {Chen}}, \bibinfo {author}
  {\bibfnamefont {B.}~\bibnamefont {Hirsch}}, \bibinfo {author} {\bibfnamefont
  {A.}~\bibnamefont {Volodin}}, \bibinfo {author} {\bibfnamefont
  {Z.}~\bibnamefont {Li}}, \bibinfo {author} {\bibfnamefont {F.}~\bibnamefont
  {Song}}, \bibinfo {author} {\bibfnamefont {J.~W.}\ \bibnamefont {Seo}},
  \bibinfo {author} {\bibfnamefont {S.~D.}\ \bibnamefont {Feyter}}, \bibinfo
  {author} {\bibfnamefont {K.}~\bibnamefont {Schouteden}}, \ and\ \bibinfo
  {author} {\bibfnamefont {C.~V.}\ \bibnamefont {Haesendonck}},\ }\bibfield
  {title} {\enquote {\bibinfo {title} {Scanning probe microscopy induced
  surface modifications of the topological insulator {Bi2Te3} in different
  environments},}\ }\href {\doibase 10.1088/1361-6528/aa7c28} {\bibfield
  {journal} {\bibinfo  {journal} {Nanotechnology}\ }\textbf {\bibinfo {volume}
  {28}},\ \bibinfo {pages} {335706} (\bibinfo {year} {2017})}\BibitemShut
  {NoStop}%
\bibitem [{\citenamefont {Music}\ \emph {et~al.}(2017)\citenamefont {Music},
  \citenamefont {Chang}, \citenamefont {Schmidt}, \citenamefont {Braun},
  \citenamefont {Heller}, \citenamefont {Hermsen}, \citenamefont {Pöllmann},
  \citenamefont {Schulzendorff},\ and\ \citenamefont
  {Wagner}}]{BiTe3_Music_2017}%
  \BibitemOpen
  \bibfield  {author} {\bibinfo {author} {\bibfnamefont {D.}~\bibnamefont
  {Music}}, \bibinfo {author} {\bibfnamefont {K.}~\bibnamefont {Chang}},
  \bibinfo {author} {\bibfnamefont {P.}~\bibnamefont {Schmidt}}, \bibinfo
  {author} {\bibfnamefont {F.~N.}\ \bibnamefont {Braun}}, \bibinfo {author}
  {\bibfnamefont {M.}~\bibnamefont {Heller}}, \bibinfo {author} {\bibfnamefont
  {S.}~\bibnamefont {Hermsen}}, \bibinfo {author} {\bibfnamefont {P.~J.}\
  \bibnamefont {Pöllmann}}, \bibinfo {author} {\bibfnamefont {T.}~\bibnamefont
  {Schulzendorff}}, \ and\ \bibinfo {author} {\bibfnamefont {C.}~\bibnamefont
  {Wagner}},\ }\bibfield  {title} {\enquote {\bibinfo {title} {On atomic
  mechanisms governing the oxidation of {Bi2Te3}},}\ }\href {\doibase
  10.1088/1361-648x/aa945f} {\bibfield  {journal} {\bibinfo  {journal} {Journal
  of Physics: Condensed Matter}\ }\textbf {\bibinfo {volume} {29}},\ \bibinfo
  {pages} {485705} (\bibinfo {year} {2017})}\BibitemShut {NoStop}%
\bibitem [{\citenamefont {Kong}\ \emph {et~al.}(2011)\citenamefont {Kong},
  \citenamefont {Cha}, \citenamefont {Lai}, \citenamefont {Peng}, \citenamefont
  {Analytis}, \citenamefont {Meister}, \citenamefont {Chen}, \citenamefont
  {Zhang}, \citenamefont {Fisher}, \citenamefont {Shen},\ and\ \citenamefont
  {Cui}}]{BiSe1_Kong}%
  \BibitemOpen
  \bibfield  {author} {\bibinfo {author} {\bibfnamefont {D.}~\bibnamefont
  {Kong}}, \bibinfo {author} {\bibfnamefont {J.~J.}\ \bibnamefont {Cha}},
  \bibinfo {author} {\bibfnamefont {K.}~\bibnamefont {Lai}}, \bibinfo {author}
  {\bibfnamefont {H.}~\bibnamefont {Peng}}, \bibinfo {author} {\bibfnamefont
  {J.~G.}\ \bibnamefont {Analytis}}, \bibinfo {author} {\bibfnamefont
  {S.}~\bibnamefont {Meister}}, \bibinfo {author} {\bibfnamefont
  {Y.}~\bibnamefont {Chen}}, \bibinfo {author} {\bibfnamefont {H.-J.}\
  \bibnamefont {Zhang}}, \bibinfo {author} {\bibfnamefont {I.~R.}\ \bibnamefont
  {Fisher}}, \bibinfo {author} {\bibfnamefont {Z.-X.}\ \bibnamefont {Shen}}, \
  and\ \bibinfo {author} {\bibfnamefont {Y.}~\bibnamefont {Cui}},\ }\bibfield
  {title} {\enquote {\bibinfo {title} {Rapid surface oxidation as a source of
  surface degradation factor for {Bi2Se3}},}\ }\href {\doibase
  10.1021/nn200556h} {\bibfield  {journal} {\bibinfo  {journal} {ACS Nano}\
  }\textbf {\bibinfo {volume} {5}},\ \bibinfo {pages} {4698--4703} (\bibinfo
  {year} {2011})}\BibitemShut {NoStop}%
\bibitem [{\citenamefont {Edmonds}\ \emph {et~al.}(2014)\citenamefont
  {Edmonds}, \citenamefont {Hellerstedt}, \citenamefont {Tadich}, \citenamefont
  {Schenk}, \citenamefont {O’Donnell}, \citenamefont {Tosado}, \citenamefont
  {Butch}, \citenamefont {Syers}, \citenamefont {Paglione},\ and\ \citenamefont
  {Fuhrer}}]{BiSe2_Edmonds}%
  \BibitemOpen
  \bibfield  {author} {\bibinfo {author} {\bibfnamefont {M.~T.}\ \bibnamefont
  {Edmonds}}, \bibinfo {author} {\bibfnamefont {J.~T.}\ \bibnamefont
  {Hellerstedt}}, \bibinfo {author} {\bibfnamefont {A.}~\bibnamefont {Tadich}},
  \bibinfo {author} {\bibfnamefont {A.}~\bibnamefont {Schenk}}, \bibinfo
  {author} {\bibfnamefont {K.~M.}\ \bibnamefont {O’Donnell}}, \bibinfo
  {author} {\bibfnamefont {J.}~\bibnamefont {Tosado}}, \bibinfo {author}
  {\bibfnamefont {N.~P.}\ \bibnamefont {Butch}}, \bibinfo {author}
  {\bibfnamefont {P.}~\bibnamefont {Syers}}, \bibinfo {author} {\bibfnamefont
  {J.}~\bibnamefont {Paglione}}, \ and\ \bibinfo {author} {\bibfnamefont
  {M.~S.}\ \bibnamefont {Fuhrer}},\ }\bibfield  {title} {\enquote {\bibinfo
  {title} {Stability and surface reconstruction of topological insulator
  {Bi2Se3} on exposure to atmosphere},}\ }\href {\doibase 10.1021/jp506089b}
  {\bibfield  {journal} {\bibinfo  {journal} {The Journal of Physical Chemistry
  C}\ }\textbf {\bibinfo {volume} {118}},\ \bibinfo {pages} {20413--20419}
  (\bibinfo {year} {2014})}\BibitemShut {NoStop}%
\bibitem [{\citenamefont {Petaccia}\ \emph {et~al.}(2009)\citenamefont
  {Petaccia}, \citenamefont {Vilmercati}, \citenamefont {Gorovikov},
  \citenamefont {Barnaba}, \citenamefont {Bianco}, \citenamefont {Cocco},
  \citenamefont {Masciovecchio},\ and\ \citenamefont
  {Goldoni}}]{PETACCIA_Elettra2009780}%
  \BibitemOpen
  \bibfield  {author} {\bibinfo {author} {\bibfnamefont {L.}~\bibnamefont
  {Petaccia}}, \bibinfo {author} {\bibfnamefont {P.}~\bibnamefont
  {Vilmercati}}, \bibinfo {author} {\bibfnamefont {S.}~\bibnamefont
  {Gorovikov}}, \bibinfo {author} {\bibfnamefont {M.}~\bibnamefont {Barnaba}},
  \bibinfo {author} {\bibfnamefont {A.}~\bibnamefont {Bianco}}, \bibinfo
  {author} {\bibfnamefont {D.}~\bibnamefont {Cocco}}, \bibinfo {author}
  {\bibfnamefont {C.}~\bibnamefont {Masciovecchio}}, \ and\ \bibinfo {author}
  {\bibfnamefont {A.}~\bibnamefont {Goldoni}},\ }\bibfield  {title} {\enquote
  {\bibinfo {title} {{BaD ElPh}: A 4m normal-incidence monochromator beamline
  at elettra},}\ }\href {\doibase https://doi.org/10.1016/j.nima.2009.05.001}
  {\bibfield  {journal} {\bibinfo  {journal} {Nuclear Instruments and Methods
  in Physics Research Section A: Accelerators, Spectrometers, Detectors and
  Associated Equipment}\ }\textbf {\bibinfo {volume} {606}},\ \bibinfo {pages}
  {780--784} (\bibinfo {year} {2009})}\BibitemShut {NoStop}%
\bibitem [{\citenamefont {K\"ufner}, \citenamefont {Fitzner},\ and\
  \citenamefont {Bechstedt}(2014)}]{Kufner2014AlhpaSn-topVS}%
  \BibitemOpen
  \bibfield  {author} {\bibinfo {author} {\bibfnamefont {S.}~\bibnamefont
  {K\"ufner}}, \bibinfo {author} {\bibfnamefont {M.}~\bibnamefont {Fitzner}}, \
  and\ \bibinfo {author} {\bibfnamefont {F.}~\bibnamefont {Bechstedt}},\
  }\bibfield  {title} {\enquote {\bibinfo {title} {Topological
  $\ensuremath{\alpha}$-{Sn} surface states versus film thickness and
  strain},}\ }\href {\doibase 10.1103/PhysRevB.90.125312} {\bibfield  {journal}
  {\bibinfo  {journal} {Phys. Rev. B}\ }\textbf {\bibinfo {volume} {90}},\
  \bibinfo {pages} {125312} (\bibinfo {year} {2014})}\BibitemShut {NoStop}%
\bibitem [{\citenamefont {Rojas-S\'anchez}\ \emph {et~al.}(2016)\citenamefont
  {Rojas-S\'anchez}, \citenamefont {Oyarz\'un}, \citenamefont {Fu},
  \citenamefont {Marty}, \citenamefont {Vergnaud}, \citenamefont {Gambarelli},
  \citenamefont {Vila}, \citenamefont {Jamet}, \citenamefont {Ohtsubo},
  \citenamefont {Taleb-Ibrahimi}, \citenamefont {Le~F\`evre}, \citenamefont
  {Bertran}, \citenamefont {Reyren}, \citenamefont {George},\ and\
  \citenamefont {Fert}}]{Rojas_PhysRevLett.116.096602}%
  \BibitemOpen
  \bibfield  {author} {\bibinfo {author} {\bibfnamefont {J.-C.}\ \bibnamefont
  {Rojas-S\'anchez}}, \bibinfo {author} {\bibfnamefont {S.}~\bibnamefont
  {Oyarz\'un}}, \bibinfo {author} {\bibfnamefont {Y.}~\bibnamefont {Fu}},
  \bibinfo {author} {\bibfnamefont {A.}~\bibnamefont {Marty}}, \bibinfo
  {author} {\bibfnamefont {C.}~\bibnamefont {Vergnaud}}, \bibinfo {author}
  {\bibfnamefont {S.}~\bibnamefont {Gambarelli}}, \bibinfo {author}
  {\bibfnamefont {L.}~\bibnamefont {Vila}}, \bibinfo {author} {\bibfnamefont
  {M.}~\bibnamefont {Jamet}}, \bibinfo {author} {\bibfnamefont
  {Y.}~\bibnamefont {Ohtsubo}}, \bibinfo {author} {\bibfnamefont
  {A.}~\bibnamefont {Taleb-Ibrahimi}}, \bibinfo {author} {\bibfnamefont
  {P.}~\bibnamefont {Le~F\`evre}}, \bibinfo {author} {\bibfnamefont
  {F.}~\bibnamefont {Bertran}}, \bibinfo {author} {\bibfnamefont
  {N.}~\bibnamefont {Reyren}}, \bibinfo {author} {\bibfnamefont {J.-M.}\
  \bibnamefont {George}}, \ and\ \bibinfo {author} {\bibfnamefont
  {A.}~\bibnamefont {Fert}},\ }\bibfield  {title} {\enquote {\bibinfo {title}
  {Spin to charge conversion at room temperature by spin pumping into a new
  type of topological insulator: $\ensuremath{\alpha}$-{Sn} films},}\ }\href
  {\doibase 10.1103/PhysRevLett.116.096602} {\bibfield  {journal} {\bibinfo
  {journal} {Phys. Rev. Lett.}\ }\textbf {\bibinfo {volume} {116}},\ \bibinfo
  {pages} {096602} (\bibinfo {year} {2016})}\BibitemShut {NoStop}%
\bibitem [{\citenamefont {Eelbo}\ \emph {et~al.}(2013)\citenamefont {Eelbo},
  \citenamefont {Sikora}, \citenamefont {Bihlmayer}, \citenamefont
  {Dobrza{\'{n}}ski}, \citenamefont {Koz{\l}owski}, \citenamefont
  {Miotkowski},\ and\ \citenamefont {Wiesendanger}}]{BiSe1_Eelbo_2013STS}%
  \BibitemOpen
  \bibfield  {author} {\bibinfo {author} {\bibfnamefont {T.}~\bibnamefont
  {Eelbo}}, \bibinfo {author} {\bibfnamefont {M.}~\bibnamefont {Sikora}},
  \bibinfo {author} {\bibfnamefont {G.}~\bibnamefont {Bihlmayer}}, \bibinfo
  {author} {\bibfnamefont {M.}~\bibnamefont {Dobrza{\'{n}}ski}}, \bibinfo
  {author} {\bibfnamefont {A.}~\bibnamefont {Koz{\l}owski}}, \bibinfo {author}
  {\bibfnamefont {I.}~\bibnamefont {Miotkowski}}, \ and\ \bibinfo {author}
  {\bibfnamefont {R.}~\bibnamefont {Wiesendanger}},\ }\bibfield  {title}
  {\enquote {\bibinfo {title} {Co atoms on {Bi2Se3} revealing a coverage
  dependent spin reorientation transition},}\ }\href {\doibase
  10.1088/1367-2630/15/11/113026} {\bibfield  {journal} {\bibinfo  {journal}
  {New Journal of Physics}\ }\textbf {\bibinfo {volume} {15}},\ \bibinfo
  {pages} {113026} (\bibinfo {year} {2013})}\BibitemShut {NoStop}%
\bibitem [{\citenamefont {Fedotov}\ and\ \citenamefont
  {Zaitsev-Zotov}(2019)}]{BiSe2_Fedotov_doi:10.1002/pssr.201800617}%
  \BibitemOpen
  \bibfield  {author} {\bibinfo {author} {\bibfnamefont {N.}~\bibnamefont
  {Fedotov}}\ and\ \bibinfo {author} {\bibfnamefont {S.}~\bibnamefont
  {Zaitsev-Zotov}},\ }\bibfield  {title} {\enquote {\bibinfo {title}
  {Experimental observation of bound states of {2D} dirac electrons at surface
  steps of the topological insulator {Bi2Se3}},}\ }\href {\doibase
  10.1002/pssr.201800617} {\bibfield  {journal} {\bibinfo  {journal} {physica
  status solidi (RRL) – Rapid Research Letters}\ }\textbf {\bibinfo {volume}
  {13}},\ \bibinfo {pages} {1800617} (\bibinfo {year} {2019})}\BibitemShut
  {NoStop}%
\bibitem [{\citenamefont {Gunnlaugsson}(2016)}]{Gunnlaugsson2016}%
  \BibitemOpen
  \bibfield  {author} {\bibinfo {author} {\bibfnamefont {H.~P.}\ \bibnamefont
  {Gunnlaugsson}},\ }\bibfield  {title} {\enquote {\bibinfo {title}
  {Spreadsheet based analysis of {M\"{o}ssbauer} spectra},}\ }\href {\doibase
  10.1007/s10751-016-1271-z} {\bibfield  {journal} {\bibinfo  {journal}
  {Hyperfine Interactions}\ }\textbf {\bibinfo {volume} {237}},\ \bibinfo
  {pages} {79} (\bibinfo {year} {2016})}\BibitemShut {NoStop}%
\end{thebibliography}%


\begin{thebibliography}{6}%
\makeatletter
\providecommand \@ifxundefined [1]{%
 \@ifx{#1\undefined}
}%
\providecommand \@ifnum [1]{%
 \ifnum #1\expandafter \@firstoftwo
 \else \expandafter \@secondoftwo
 \fi
}%
\providecommand \@ifx [1]{%
 \ifx #1\expandafter \@firstoftwo
 \else \expandafter \@secondoftwo
 \fi
}%
\providecommand \natexlab [1]{#1}%
\providecommand \enquote  [1]{``#1''}%
\providecommand \bibnamefont  [1]{#1}%
\providecommand \bibfnamefont [1]{#1}%
\providecommand \citenamefont [1]{#1}%
\providecommand \href@noop [0]{\@secondoftwo}%
\providecommand \href [0]{\begingroup \@sanitize@url \@href}%
\providecommand \@href[1]{\@@startlink{#1}\@@href}%
\providecommand \@@href[1]{\endgroup#1\@@endlink}%
\providecommand \@sanitize@url [0]{\catcode `\\12\catcode `\$12\catcode
  `\&12\catcode `\#12\catcode `\^12\catcode `\_12\catcode `\%12\relax}%
\providecommand \@@startlink[1]{}%
\providecommand \@@endlink[0]{}%
\providecommand \url  [0]{\begingroup\@sanitize@url \@url }%
\providecommand \@url [1]{\endgroup\@href {#1}{\urlprefix }}%
\providecommand \urlprefix  [0]{URL }%
\providecommand \Eprint [0]{\href }%
\providecommand \doibase [0]{http://dx.doi.org/}%
\providecommand \selectlanguage [0]{\@gobble}%
\providecommand \bibinfo  [0]{\@secondoftwo}%
\providecommand \bibfield  [0]{\@secondoftwo}%
\providecommand \translation [1]{[#1]}%
\providecommand \BibitemOpen [0]{}%
\providecommand \bibitemStop [0]{}%
\providecommand \bibitemNoStop [0]{.\EOS\space}%
\providecommand \EOS [0]{\spacefactor3000\relax}%
\providecommand \BibitemShut  [1]{\csname bibitem#1\endcsname}%
\let\auto@bib@innerbib\@empty
\bibitem [{\citenamefont {Farrow}\ \emph {et~al.}(1981)\citenamefont {Farrow},
  \citenamefont {Robertson}, \citenamefont {Williams}, \citenamefont {Cullis},
  \citenamefont {Jones}, \citenamefont {Young},\ and\ \citenamefont
  {Dennis}}]{FARROW1981507}%
  \BibitemOpen
  \bibfield  {author} {\bibinfo {author} {\bibfnamefont {R.}~\bibnamefont
  {Farrow}}, \bibinfo {author} {\bibfnamefont {D.}~\bibnamefont {Robertson}},
  \bibinfo {author} {\bibfnamefont {G.}~\bibnamefont {Williams}}, \bibinfo
  {author} {\bibfnamefont {A.}~\bibnamefont {Cullis}}, \bibinfo {author}
  {\bibfnamefont {G.}~\bibnamefont {Jones}}, \bibinfo {author} {\bibfnamefont
  {I.}~\bibnamefont {Young}}, \ and\ \bibinfo {author} {\bibfnamefont
  {P.}~\bibnamefont {Dennis}},\ }\bibfield  {title} {\enquote {\bibinfo {title}
  {The growth of metastable, heteroepitaxial films of \mbox{$\alpha$-Sn} by
  metal beam epitaxy},}\ }\href {\doibase
  https://doi.org/10.1016/0022-0248(81)90506-6} {\bibfield  {journal} {\bibinfo
   {journal} {Journal of Crystal Growth}\ }\textbf {\bibinfo {volume} {54}},\
  \bibinfo {pages} {507--518} (\bibinfo {year} {1981})}\BibitemShut {NoStop}%
\bibitem [{\citenamefont {Ueda}\ \emph {et~al.}(1991)\citenamefont {Ueda},
  \citenamefont {Nakayama}, \citenamefont {Sekine},\ and\ \citenamefont
  {Fujita}}]{UEDA1991AUGER}%
  \BibitemOpen
  \bibfield  {author} {\bibinfo {author} {\bibfnamefont {K.}~\bibnamefont
  {Ueda}}, \bibinfo {author} {\bibfnamefont {H.}~\bibnamefont {Nakayama}},
  \bibinfo {author} {\bibfnamefont {M.}~\bibnamefont {Sekine}}, \ and\ \bibinfo
  {author} {\bibfnamefont {H.}~\bibnamefont {Fujita}},\ }\bibfield  {title}
  {\enquote {\bibinfo {title} {Auger valence electron spectroscopy of a
  structural phase transformation in metastable alpha-{Sn} grown on {InSb}
  (001)},}\ }\href {\doibase https://doi.org/10.1016/0042-207X(91)90935-C}
  {\bibfield  {journal} {\bibinfo  {journal} {Vacuum}\ }\textbf {\bibinfo
  {volume} {42}},\ \bibinfo {pages} {547} (\bibinfo {year} {1991})}\BibitemShut
  {NoStop}%
\bibitem [{\citenamefont {Betti}\ \emph {et~al.}(2002)\citenamefont {Betti},
  \citenamefont {Magnano}, \citenamefont {Sancrotti}, \citenamefont {Borgatti},
  \citenamefont {Felici}, \citenamefont {Mariani},\ and\ \citenamefont
  {Sauvage-Simkin}}]{BETTI2002335}%
  \BibitemOpen
  \bibfield  {author} {\bibinfo {author} {\bibfnamefont {M.~G.}\ \bibnamefont
  {Betti}}, \bibinfo {author} {\bibfnamefont {E.}~\bibnamefont {Magnano}},
  \bibinfo {author} {\bibfnamefont {M.}~\bibnamefont {Sancrotti}}, \bibinfo
  {author} {\bibfnamefont {F.}~\bibnamefont {Borgatti}}, \bibinfo {author}
  {\bibfnamefont {R.}~\bibnamefont {Felici}}, \bibinfo {author} {\bibfnamefont
  {C.}~\bibnamefont {Mariani}}, \ and\ \bibinfo {author} {\bibfnamefont
  {M.}~\bibnamefont {Sauvage-Simkin}},\ }\bibfield  {title} {\enquote {\bibinfo
  {title} {Growth morphology of ($1\times2$) \mbox{$\alpha$-Sn(100)}: a surface
  diffraction study},}\ }\href {\doibase
  https://doi.org/10.1016/S0039-6028(02)01267-0} {\bibfield  {journal}
  {\bibinfo  {journal} {Surface Science}\ }\textbf {\bibinfo {volume}
  {507-510}},\ \bibinfo {pages} {335--339} (\bibinfo {year}
  {2002})}\BibitemShut {NoStop}%
\bibitem [{\citenamefont {Magnano}\ \emph {et~al.}(2002)\citenamefont
  {Magnano}, \citenamefont {Cepek}, \citenamefont {Gardonio}, \citenamefont
  {Allieri}, \citenamefont {Baek}, \citenamefont {Vescovo}, \citenamefont
  {Roca}, \citenamefont {Avila}, \citenamefont {Betti}, \citenamefont
  {Mariani},\ and\ \citenamefont {Sancrotti}}]{MAGNANO200229}%
  \BibitemOpen
  \bibfield  {author} {\bibinfo {author} {\bibfnamefont {E.}~\bibnamefont
  {Magnano}}, \bibinfo {author} {\bibfnamefont {C.}~\bibnamefont {Cepek}},
  \bibinfo {author} {\bibfnamefont {S.}~\bibnamefont {Gardonio}}, \bibinfo
  {author} {\bibfnamefont {B.}~\bibnamefont {Allieri}}, \bibinfo {author}
  {\bibfnamefont {I.}~\bibnamefont {Baek}}, \bibinfo {author} {\bibfnamefont
  {E.}~\bibnamefont {Vescovo}}, \bibinfo {author} {\bibfnamefont
  {L.}~\bibnamefont {Roca}}, \bibinfo {author} {\bibfnamefont {J.}~\bibnamefont
  {Avila}}, \bibinfo {author} {\bibfnamefont {M.~G.}\ \bibnamefont {Betti}},
  \bibinfo {author} {\bibfnamefont {C.}~\bibnamefont {Mariani}}, \ and\
  \bibinfo {author} {\bibfnamefont {M.}~\bibnamefont {Sancrotti}},\ }\bibfield
  {title} {\enquote {\bibinfo {title} {Sn on {InSb}(100)–c($2\times8$):
  growth morphology and electronic structure},}\ }\href {\doibase
  https://doi.org/10.1016/S0368-2048(02)00169-X} {\bibfield  {journal}
  {\bibinfo  {journal} {J Electron Spectrosc}\ }\textbf {\bibinfo {volume}
  {127}},\ \bibinfo {pages} {29} (\bibinfo {year} {2002})}\BibitemShut
  {NoStop}%
\bibitem [{\citenamefont {Moulder}\ and\ \citenamefont
  {Chastain}(1992)}]{XPSmoulder1992handbook}%
  \BibitemOpen
  \bibfield  {author} {\bibinfo {author} {\bibfnamefont {J.}~\bibnamefont
  {Moulder}}\ and\ \bibinfo {author} {\bibfnamefont {J.}~\bibnamefont
  {Chastain}},\ }\href@noop {} {\emph {\bibinfo {title} {Handbook of X-ray
  Photoelectron Spectroscopy: A Reference Book of Standard Spectra for
  Identification and Interpretation of {XPS} Data}}}\ (\bibinfo  {publisher}
  {Physical Electronics Division, Perkin-Elmer Corporation},\ \bibinfo {year}
  {1992})\BibitemShut {NoStop}%
\bibitem [{\citenamefont {Greenwood}\ and\ \citenamefont
  {Gibb}(1971)}]{Greenwood1971}%
  \BibitemOpen
  \bibfield  {author} {\bibinfo {author} {\bibfnamefont {N.~N.}\ \bibnamefont
  {Greenwood}}\ and\ \bibinfo {author} {\bibfnamefont {T.~C.}\ \bibnamefont
  {Gibb}},\ }\href {\doibase 10.1007/978-94-009-5697-1} {\emph {\bibinfo
  {title} {M\"{o}ssbauer Spectroscopy}}}\ (\bibinfo  {publisher} {Springer
  Netherlands},\ \bibinfo {year} {1971})\BibitemShut {NoStop}%
\end{thebibliography}%

\begin{figure}
\includegraphics[width=236.84843pt]{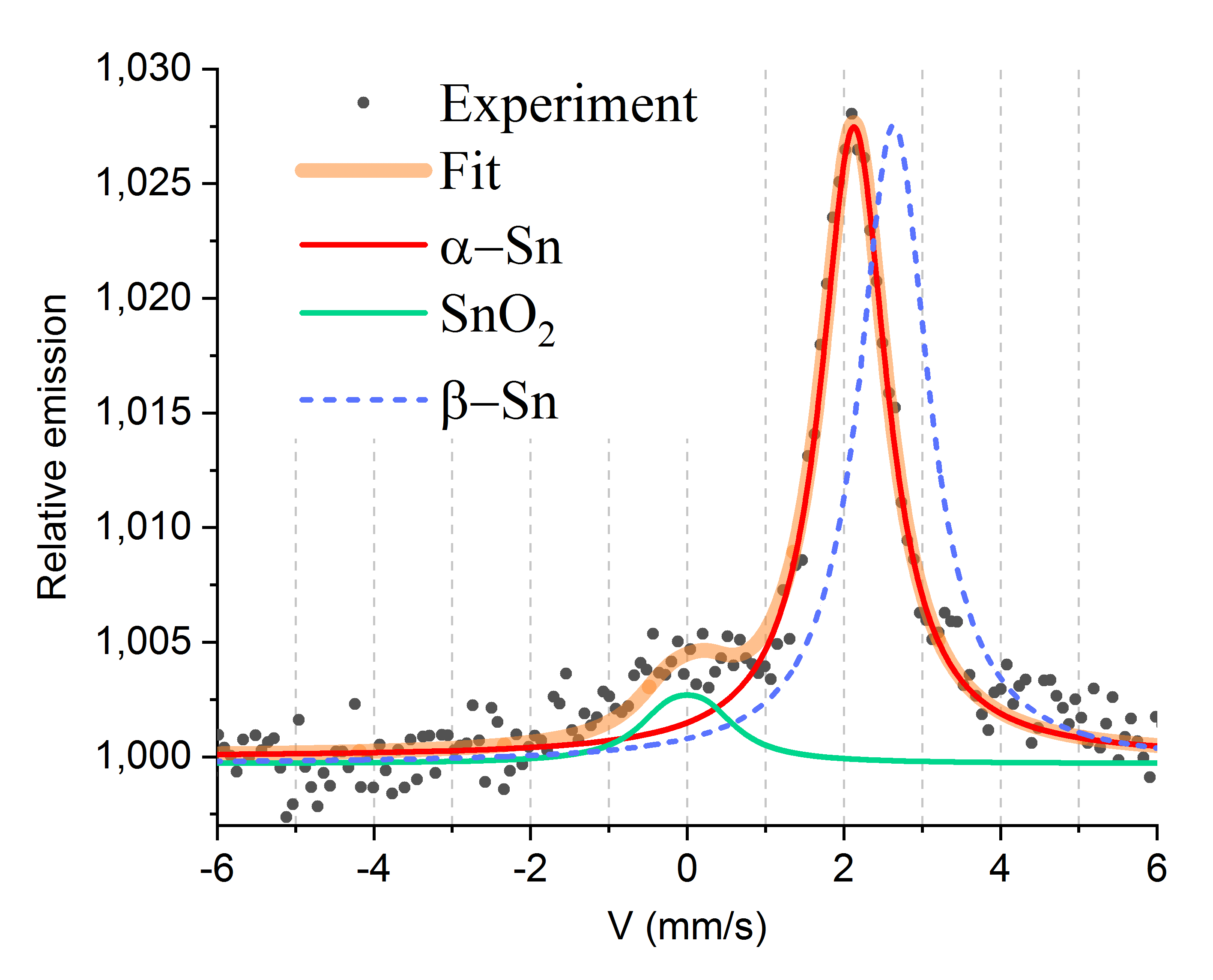}
\caption{\label{fig3} CEMS (room temperature) spectrum of a \mbox{20 nm} thick \mbox{$\alpha$-Sn} film grown at a substrate temperature of \mbox{5$^{\circ}$C}. Experimental data points (black dots) with the fit (thick orange line) are showing a dominating Mössbauer peak of \mbox{$\alpha$-Sn} (red) and a peak from SnO$_2$ (green). The added blue dashed line illustrates the position of the \mbox{$\beta$-Sn} peak as observed in the test Sn film grown on a SiO$_2$/Si substrate.}
 \end{figure}

 \begin{figure*}
 \includegraphics[width=\textwidth,height=\textheight,keepaspectratio]{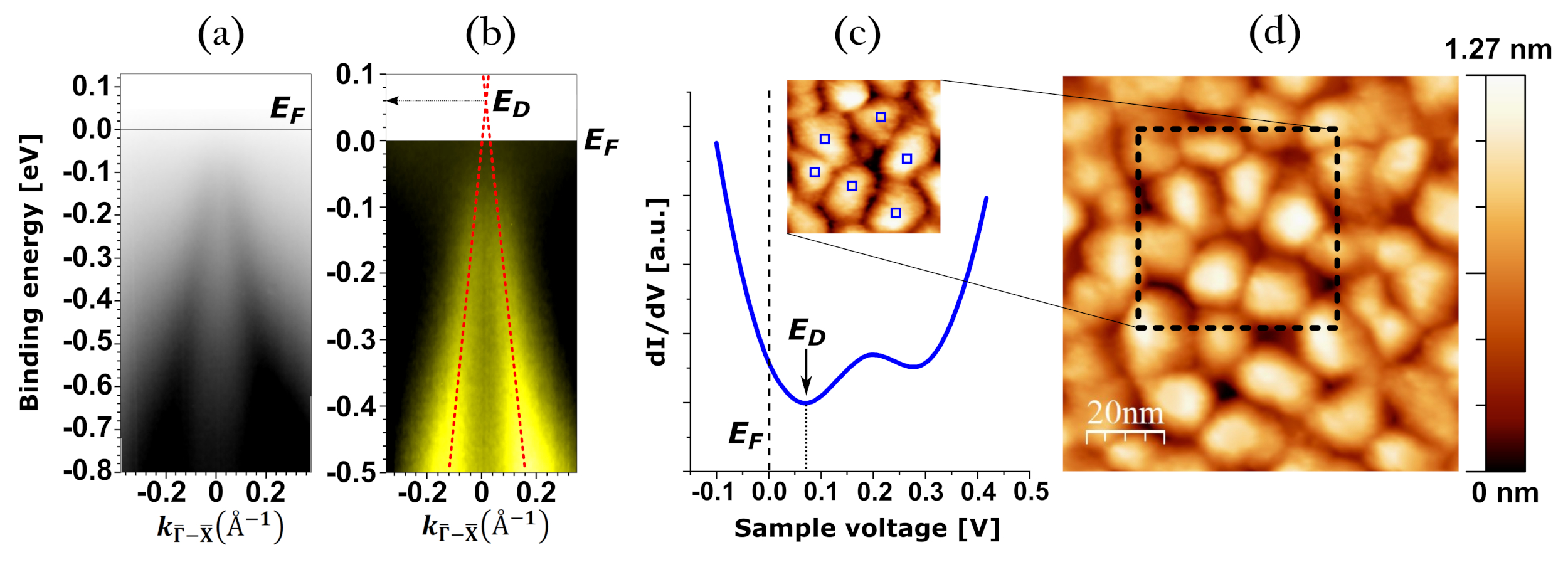}
 \caption{\label{fig4} ARPES and STS spectroscopy of a \mbox{20 nm} thick \mbox{$\alpha$-Sn} film grown at a substrate temperature of \mbox{5$^{\circ}$C}. (a) ARPES (21 eV) spectrum image measured at 77 K, along the $\overline{\Gamma}\textnormal{-}\overline{X}$ direction (originally acquired intensity plot); (b) ARPES image after subtraction of the background intensity, together with the line-profile fit; (c) STS $dI/dV$ spectrum (4.5 K) acquired on top of the Sn grains. The spectrum is area averaged \mbox{(5 nm $\times$ 5 nm blue squares)} and averaged over 6 grains. $E_D$ marks the Dirac point position above the Fermi energy ($E_F$); (d) \mbox{100 nm} $\times$ \mbox{100 nm} STM topographic image (\mbox{$U = 1$ V}, \mbox{$I = 100$ pA}) acquired at 4.5 K.}
\end{figure*}

\end{document}


\preprint{AIP/123-QED}

\title[APM19-AR-01221]{\huge Supplementary material \\
\vspace{10pt}
\large Structural and electronic properties of the pure and stable elemental\\ 3D topological Dirac semimetal $\alpha$-Sn
\vspace{10pt}}

\author{Ivan Madarevic}
\author{Umamahesh Thupakula}%
\affiliation{ 
Quantum Solid State Physics, KU Leuven, Celestijnenlaan 200D, 3001 Leuven, Belgium
}
 \author{Gertjan Lippertz}
 \affiliation{ 
Quantum Solid State Physics, KU Leuven, Celestijnenlaan 200D, 3001 Leuven, Belgium
}
\affiliation{ 
Physics Institute II, University of Cologne, Zülpicher Str. 77, 50937 Köln, Germany
}
 \author{Niels Claessens}
 \affiliation{ 
Quantum Solid State Physics, KU Leuven, Celestijnenlaan 200D, 3001 Leuven, Belgium
}
\affiliation{ 
IMEC, Kapeldreef 75, 3001 Leuven, Belgium
}
 \author{Pin-Cheng Lin}
  \author{Harsh Bana}
  \affiliation{ 
Quantum Solid State Physics, KU Leuven, Celestijnenlaan 200D, 3001 Leuven, Belgium
}%

\author{Sara Gonzalez}
\author{Giovanni Di Santo}
\author{Luca Petaccia}
\affiliation{Elettra Sincrotrone Trieste, Strada Statale 14 km 163.5, 34149 Trieste, Italy
}%
\author{Maya Narayanan Nair}
\affiliation{ 
Quantum Solid State Physics, KU Leuven, Celestijnenlaan 200D, 3001 Leuven, Belgium
}%
\affiliation{ 
CUNY Advanced Science Research Center, 85 St. Nicholas Terrace, New York, NY 10031, USA
}
\author{Lino M.C. Pereira}
\author{Chris Van Haesendonck}
\author{Margriet J. Van Bael}
\affiliation{ 
Quantum Solid State Physics, KU Leuven, Celestijnenlaan 200D, 3001 Leuven, Belgium
}%

\date{\today}

\maketitle

\section{I\MakeLowercase{n}S\MakeLowercase{b}(100) substrate preparation and characterization}

As stated in the main text, an appropriate substrate is needed to provide a seed for the growth of the alpha phase of Sn. In this case an InSb(100) substrate is used, which has a lattice constant of 6.479 \AA. Comparing this value with the lattice constant of \mbox{$\alpha$-Sn} \mbox{(6.489 \AA)}, we can assume that \mbox{0.14 \%} in-plane compressive strain is introduced in the \mbox{$\alpha$-Sn} films grown on InSb(100) substrates \cite{FARROW1981507,UEDA1991AUGER,BETTI2002335,MAGNANO200229}.

\begin{figure}[h]
\includegraphics[width=\linewidth]{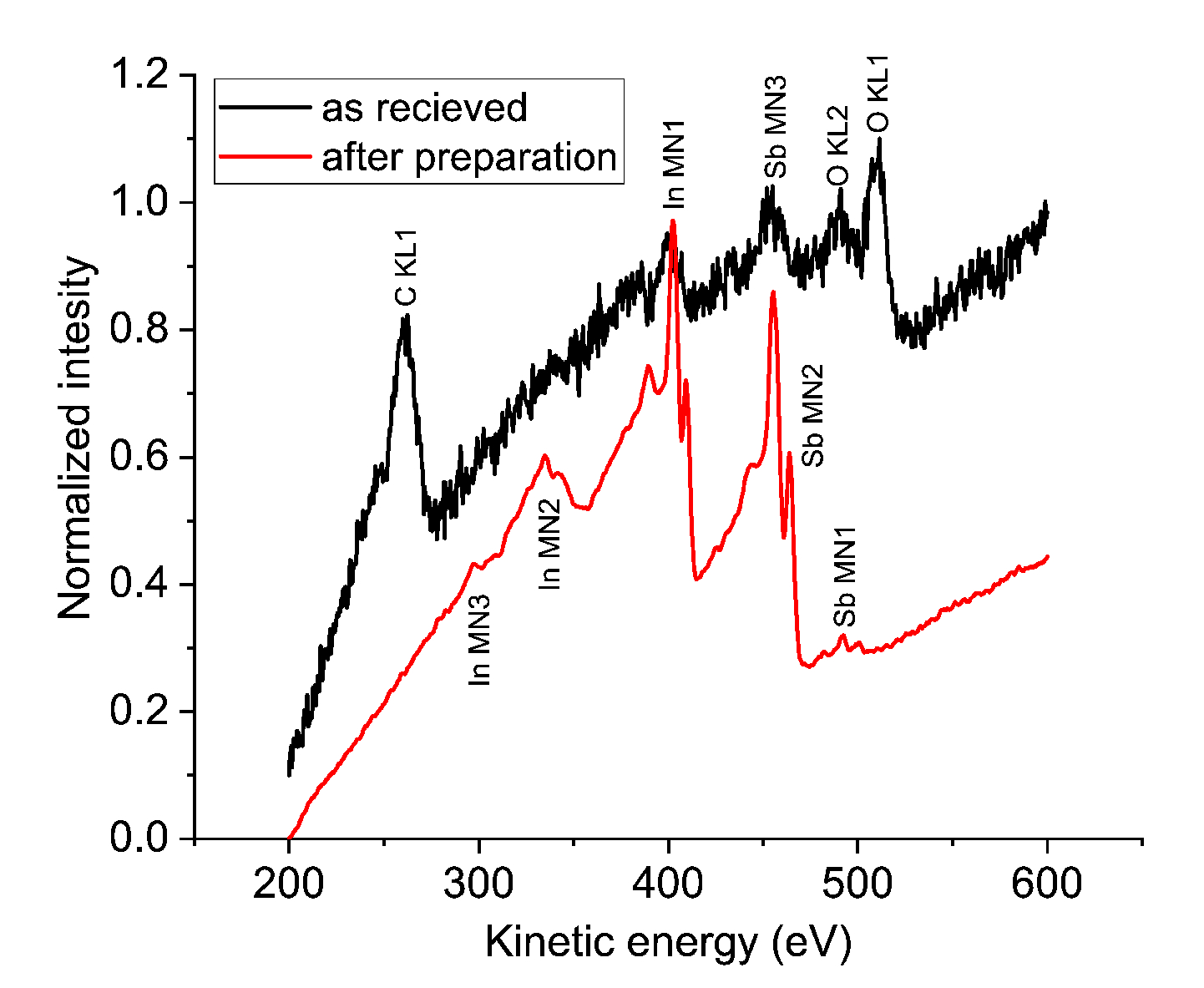}
\caption{\label{fig1s} AES spectrum of the InSb(100) substrate after preparation (red), AES spectrum of the InSb(100) as received (black) given with an offset.}
\end{figure}

InSb(100) substrates were cut to a size of approximately \mbox{$7 \times 7\; \mathrm{mm}^2$} and prepared in UHV environment by consecutive cycles of \mbox{Ar$^+$} sputtering and thermal annealing. The use of the low energy ion-beam mode (\mbox{500 eV}) allowed us to increase the sputtering time and successfully remove C and O contamination on top of the InSb(100), without introducing any irreversible damage to its surface. The higher energy ion-beam mode (1000 eV), used with the angle of incidence of 60$^{\circ}$, proved useful for the gentle polishing of the surface of the substrate. Specially defined short multi-step thermal annealing procedure (up to 300$^{\circ}$C) was used after every \mbox{Ar$^+$} sputtering treatment (\mbox{500 eV} or \mbox{1000 eV}), in order to reconstruct the surface of the substrate. Auger electron spectroscopy (AES) was systematically used before and after these cycles to monitor the chemical composition of the top layers of the substrate (Fig.~\ref{fig1s}). AES scans after the preparation show a chemically clean surface with the In/Sb stoichiometry maintained.

\begin{figure*}
\includegraphics[width=\textwidth,height=\textheight,keepaspectratio]{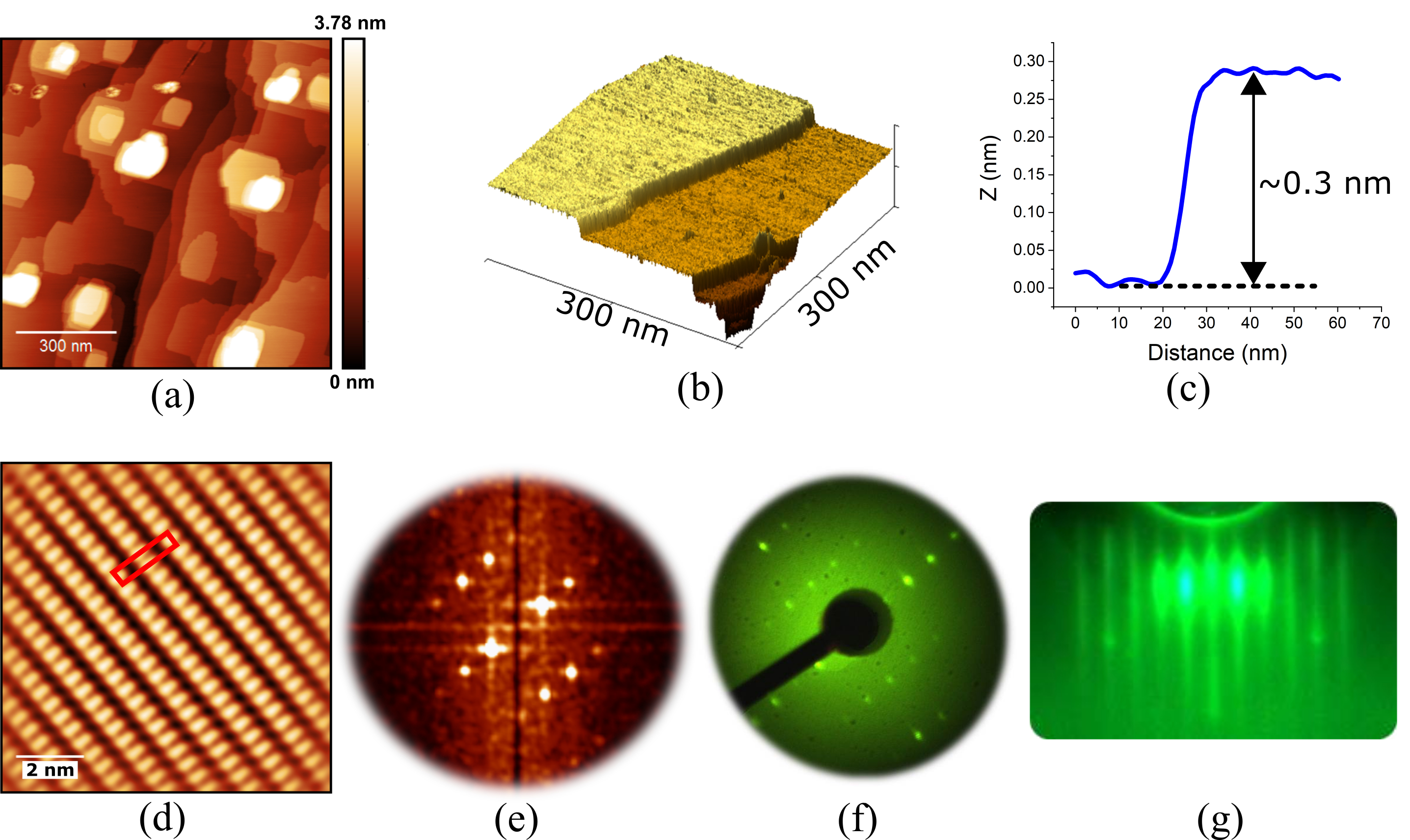}
\caption{\label{fig2s} Surface structure and morphology of the prepared InSb(100) substrate. (a) \mbox{1000 nm} $\times$ \mbox{1000 nm} STM (room temperature) topography image (\mbox{$U = 2$ V}, \mbox{$I = 100$ pA}); (b) 3D STM image (\mbox{300 nm} $\times$ \mbox{300 nm}) of the step features (\mbox{$U = 1$ V}, \mbox{$I = 150$ pA}); (c) STM line profile providing the step height, taken along the step shown in (b); (d) Atomic resolution FFT filtered STM image (\mbox{10 nm} $\times$ \mbox{10 nm}, \mbox{$U = -1.5$ V}, \mbox{$I = 70$ pA}); (e) FFT of the atomic resolution STM image; (f) and (g) show LEED (\mbox{52 eV}) image and RHEED (\mbox{10 keV}) image, respectively, revealing the \mbox{c($8 \times 2)$} surface reconstruction.}
\end{figure*}

During and after the above described in situ cleaning process, low-energy electron diffraction (LEED) and reflection high-energy electron diffraction (RHEED) were employed to monitor the surface reconstruction of the substrate. At the end of the preparation process, sharp LEED and RHEED images were acquired \mbox{(Fig.~\ref{fig2s}(f) and (g))} showing the \mbox{c($8 \times 2$)} surface reconstruction of InSb(100). RHEED images revealed good overall flatness of the InSb surface, showing streaky patterns. In order to determine the surface roughness, scanning tunneling microscopy (STM) was used. STM topography images exhibit flat terraced (100) surfaces \mbox{(Fig.~\ref{fig2s}(a) and (b))} with a step height value of \mbox{$\approx0.3$ nm}, matching the $d_{100}$ of InSb \mbox{(Fig.~\ref{fig2s}(c))}. The acquired atomic resolution STM images \mbox{(Fig.~\ref{fig2s}(d) and (e))} confirm the \mbox{c($8 \times 2$)} surface reconstruction.

\section{X-ray photoemission spectroscopy \\of the $\alpha$-S\MakeLowercase{n} films grown at 5$^{\circ}$C substrate temperature}

\begin{figure}[h]
\includegraphics[width=\linewidth]{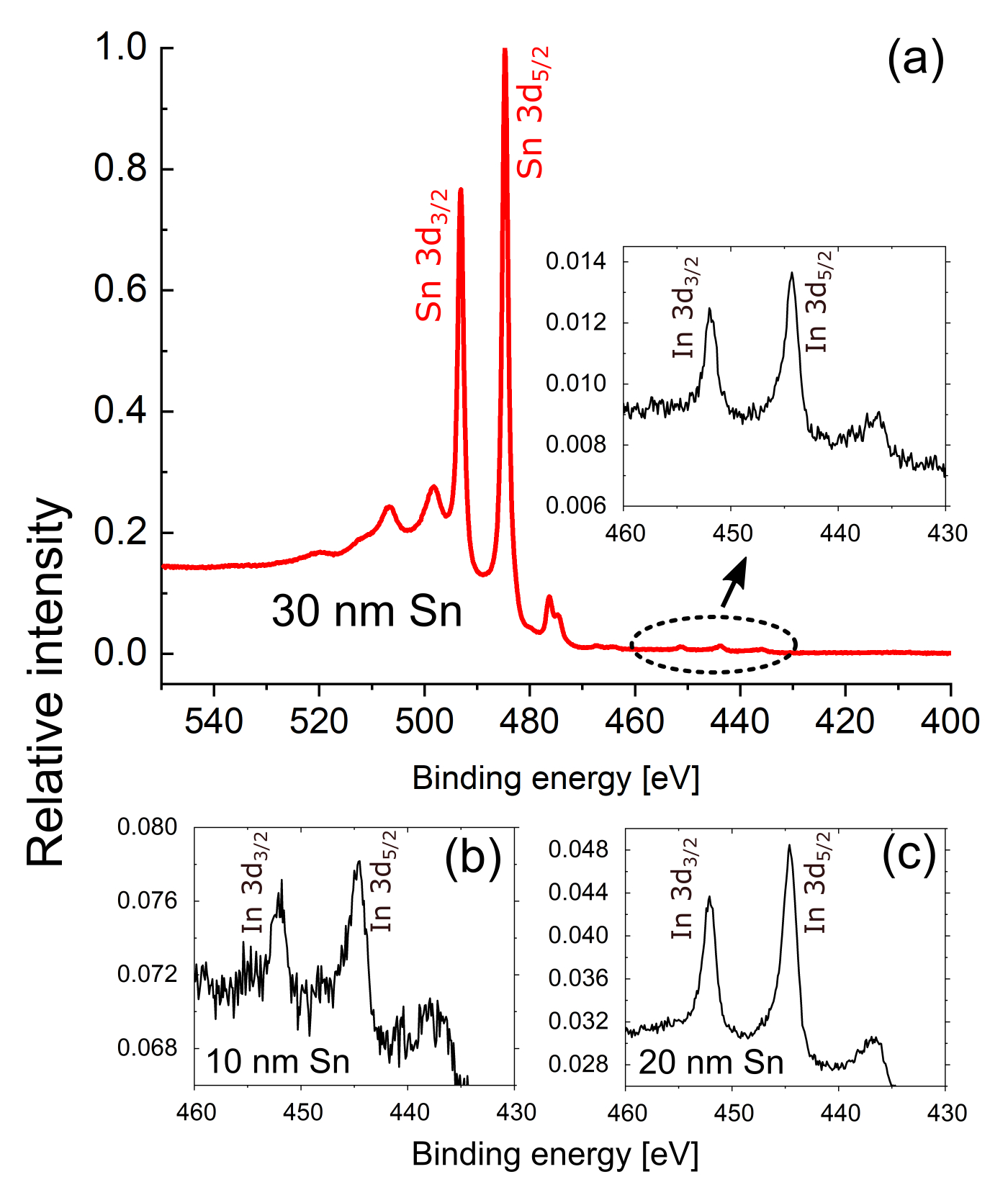}
\caption{\label{fig3s} (a) XPS (3d spectrum) of a \mbox{30 nm} thick \mbox{$\alpha$-Sn} film grown at 5$^{\circ}$C substrate temperature; XPS signal from In (3d spectra) in the cases of \mbox{10 nm} (b) and \mbox{20 nm} (c) thick \mbox{$\alpha$-Sn} films.}
\end{figure}

To further investigate the subsurface chemical composition of the \mbox{$\alpha$-Sn} samples grown at 5$^{\circ}$C substrate temperature, \mbox{X-ray} photoemission spectroscopy (XPS) was used. Analysis of the acquired XPS spectra \mbox{(Fig.~\ref{fig3s})} shows a small presence of In (\mbox{< 3 \%}, \mbox{< 2 \%} and \mbox{< 1 \%} for 10, 20 and \mbox{30 nm} thick Sn films, respectively), estimated by comparing the intensities of the $3d_{5/2}$ peaks \cite{XPSmoulder1992handbook}. Since in the case of 20 and 30 nm thick films grown at 5$^{\circ}$C substrate temperature no traces of In were detected by AES, we conclude that the XPS In signal must come either from the subsurface layers of the film, the edges of the sample or the substrate areas shaded by clips during Sn deposition.

\section{S\MakeLowercase{n} films grown on slightly heated\\ I\MakeLowercase{n}S\MakeLowercase{b}(100) substrates}

\begin{figure}[h]
\includegraphics[width=\linewidth]{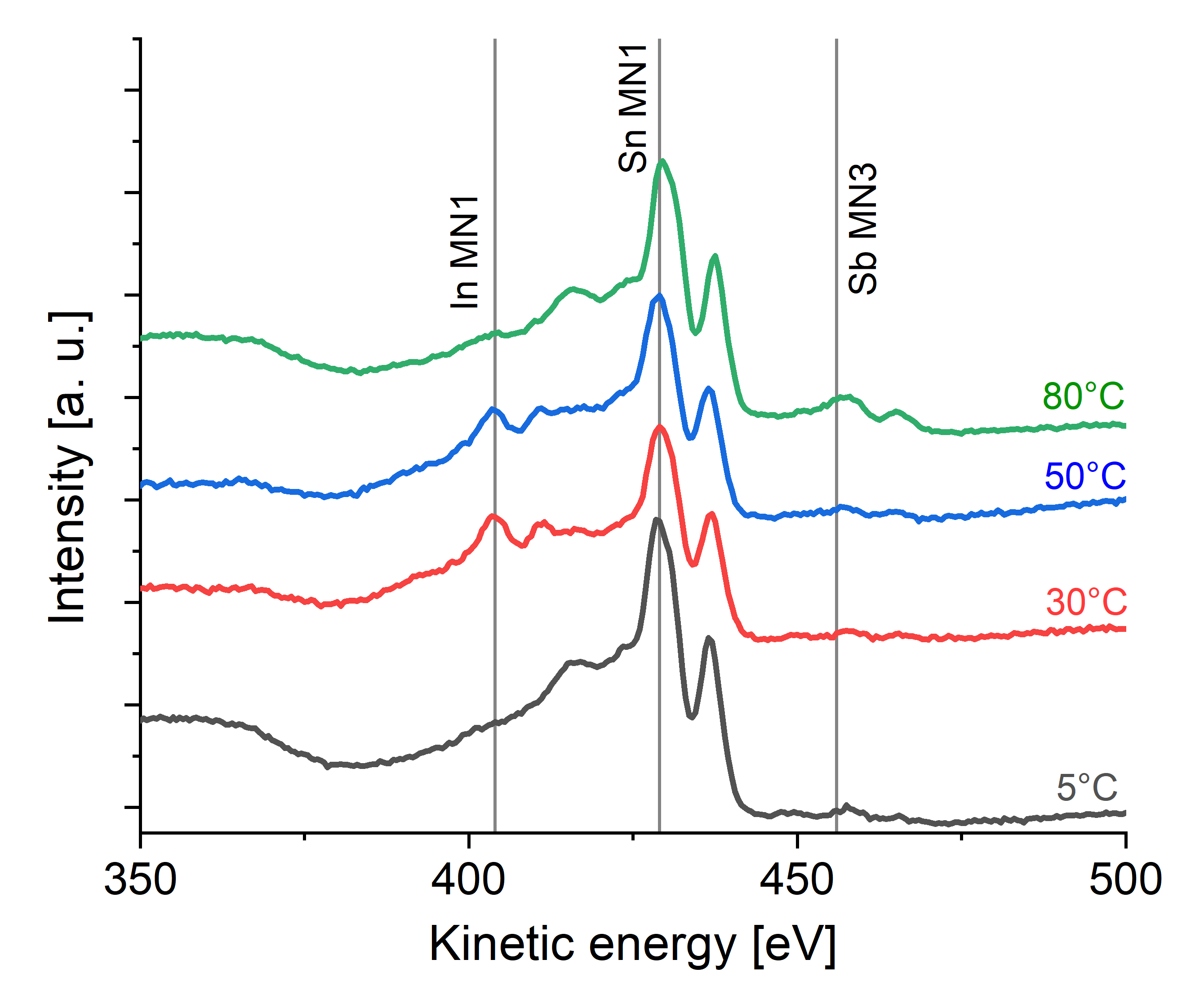}
\caption{\label{fig4s} AES spectra of 30 nm thick \mbox{$\alpha$-Sn} films grown at different InSb(100) substrate temperatures; 5$^{\circ}$C (black), 30$^{\circ}$C (red), 50$^{\circ}$C (blue) and 80$^{\circ}$C (green).}
 \end{figure}

\mbox{Figure~\ref{fig4s}} presents a comparison of the acquired AES spectra of the Sn samples grown at different temperatures of the InSb(100) substrate. Differently from the samples grown at a substrate temperature of 5$^{\circ}$C, where no In was detected by AES, its presence was evident in the AES spectra of the Sn samples grown at a substrate temperature of 30$^{\circ}$C, 50$^{\circ}$C and 80$^{\circ}$C. Interestingly, in the samples grown at 80$^{\circ}$C comparable amounts of Sb were also detected, implying that above certain temperatures infusion of Sb into the film occurs as well.

\begin{figure*}[h]
\includegraphics[width=\linewidth]{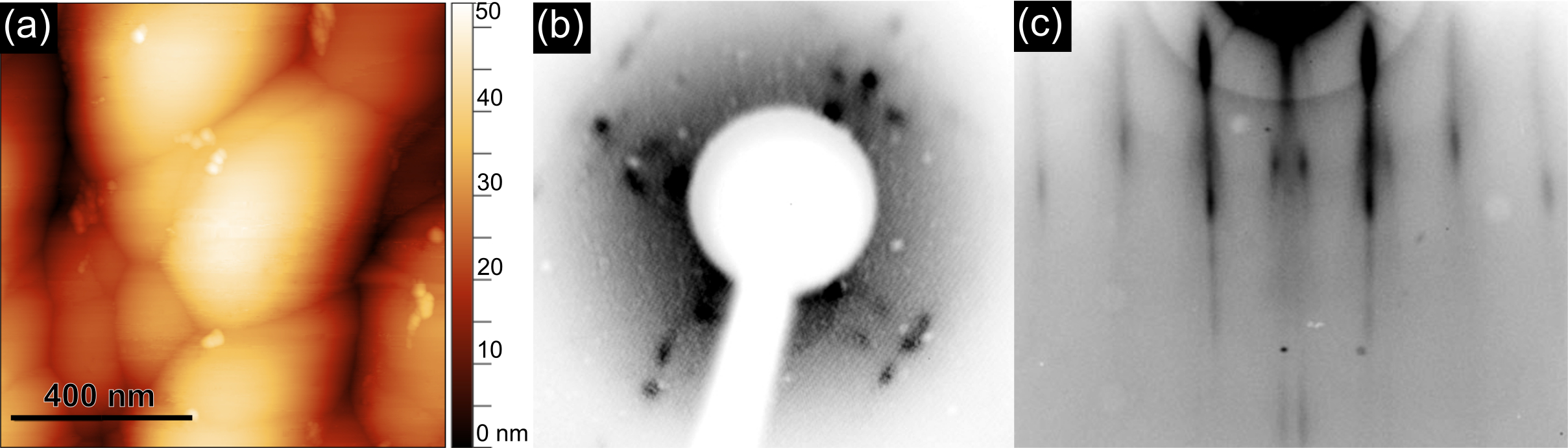}
 \caption{\label{fig5s} Room-temperature STM topography of a 30 nm thick \mbox{$\alpha$-Sn} film grown on InSb(100) at a substrate temperature of 50$^{\circ}$C. (a) \mbox{1000 nm} $\times$ \mbox{1000 nm} STM topography image (\mbox{$U = -2$ V}, \mbox{$I = 70$ pA}); (b) LEED (48 eV) and (c) RHEED (10 kV) image of a 30 nm thick \mbox{$\alpha$-Sn} film grown at 50$^{\circ}$C InSb(100) substrate temperature.}
 \end{figure*}

\mbox{Figure~\ref{fig5s}(a)} shows the room-temperature STM topography images of a Sn film grown at a substrate temperature of 50$^{\circ}$C. In contrast to the samples grown at a substrate temperature of 5$^{\circ}$C (grain-structured surface with low roughness), the surface morphology here exhibits many different features and an enhanced surface roughness (\mbox{$RMS_{1000}\sim7.8\; \mathrm{nm}$} -- \textit{root mean square roughness} taken over a $1000 \times 1000\; \mathrm{nm}^2$ area), including the appearances of wide islands.

Observed LEED and RHEED patterns (\mbox{Fig.~\ref{fig5s}(b) and (c)}) appeared in this case are distorted, compared to the ones acquired on the samples grow at 5$^{\circ}$C, making the full analysis of the images very difficult. Unfortunately, one can only choose to speculate that the acquired data is the consequence of the evenly disturbed epitaxial growth and chemical purity of the Sn film.

\section{Mössbauer spectroscopy of a reference $\beta$-S\MakeLowercase{n} film}

\begin{figure}[h]
\includegraphics[width=\linewidth]{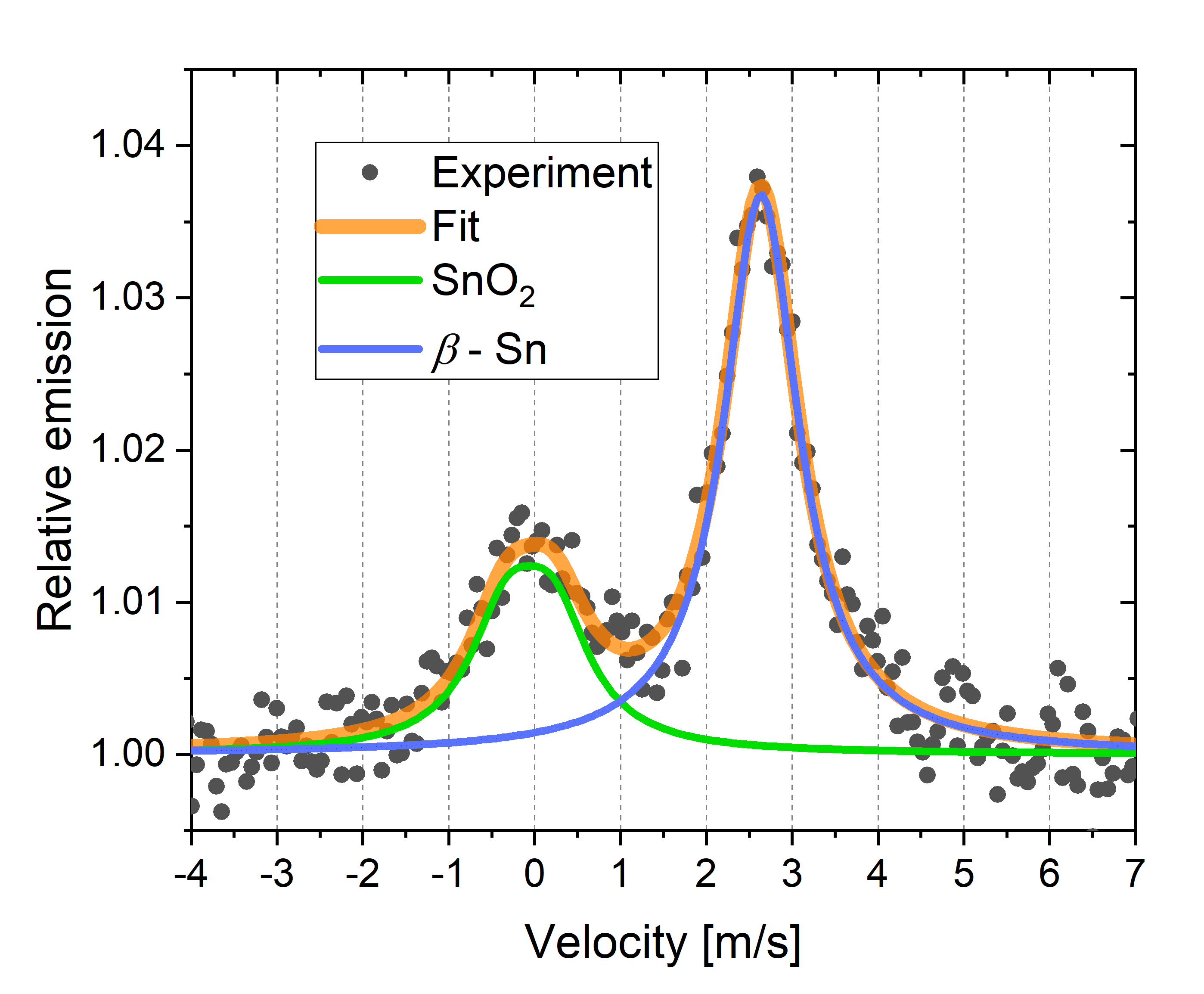}
\caption{\label{fig6s} CEMS (room temperature) spectrum of a \mbox{30 nm} thick \mbox{$\beta$-Sn} film grown on SiO$_2$/Si substrate (at 5$^{\circ}$C temperature). Experimental data points (black dots) with the fit (thick black line) reflect the presence of a dominating Mössbauer peak of \mbox{$\beta$-Sn} (blue) and a peak from SnO$_2$ (green).}
\end{figure}

\mbox{Figure~\ref{fig6s}} presents the conversion electron Mössbauer spectroscopy (CEMS) spectrum of a reference polycrystalline \mbox{30 nm} thick \mbox{$\beta$-Sn} film, grown on SiO$_2$/Si substrate. The acquired spectrum is dominated by the \mbox{$\beta$-Sn} peak at \mbox{2.64(1) mm/s}, considering that the CEMS sensitivity for $\mathrm{SnO}_2$ (whose peak is observed at \mbox{0 mm/s}) is approximately ten times larger compared to that of \mbox{$\beta$-Sn} \cite{Greenwood1971}. The peak from \mbox{$\alpha$-Sn} is completely absent in this case, as expected, since the $\mathrm{SiO}_2$/Si substrate is not suitable for the growth of \mbox{$\alpha$-Sn}.

\section{Cleaning of the exposed \mbox{$\alpha$-S\MakeLowercase{n}} by argon beam sputtering}

\begin{figure}[h]
\includegraphics[width=\linewidth]{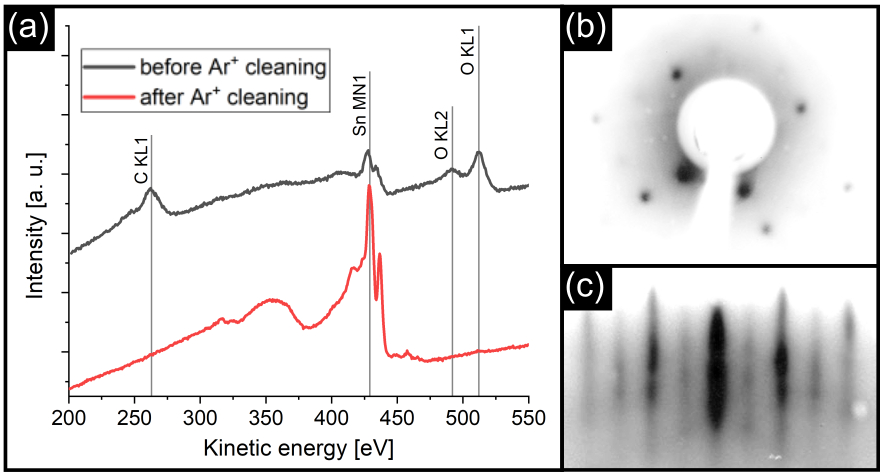}
\caption{\label{fig7s} \mbox{30 nm} \mbox{$\alpha$-Sn} film grown at a substrate temperature of \mbox{5$^{\circ}$C}, cleaned by \mbox{Ar$^+$} sputtering. (a) AES spectrum taken before (black) and after cleaning (red); (b) LEED (\mbox{48 eV}) image acquired after cleaning, showing two-domain (\mbox{2$\times$1}) pattern; (c) RHEED (\mbox{10 kV}) image after cleaning, acquired along the [110] direction, showing two-domain (\mbox{2$\times$1}) pattern.}
\end{figure}

Now we present more details about the successful \mbox{Ar$^+$} beam cleaning (after being exposed to ambient conditions) of the \mbox{$\alpha$-Sn} films grown at a substrate temperature of 5$^{\circ}$C. This cleaning procedure was done under UHV conditions prior to the angle-resolved photoemission spectroscopy (ARPES) and scanning tunneling spectroscopy (STS) measurements. By using the current of $6\times10^{-2}$ $\mu$A/mm$^2$ low-energy argon plasma \mbox{(500 eV)} during 1h, we were able to completely clean the 20 and 30 nm \mbox{$\alpha$-Sn} films which were previously kept under ambient conditions for several weeks. \mbox{Figure~\ref{fig7s}(a)} shows the comparison between the AES spectrum recorded before and after the \mbox{Ar$^+$} beam treatment. The AES spectrum recorded after this cleaning process appears identical to the one acquired just after the initial Sn film deposition.

Importantly, after this cleaning treatment we also observe a complete recovery of the two-domain ($2\times1$) surface reconstruction of \mbox{$\alpha$-Sn} \mbox{({Fig.~\ref{fig7s}(b) and (c)})}, using electron diffraction (LEED and RHEED). These findings imply remarkable stability and robustness of the sample surface against exposure to ambient conditions and subsequent \mbox{Ar$^+$} sputtering.

\section*{References}
\bibliography{aipsamp}

